\documentclass[%
 aip,
 amsmath,amssymb,
 reprint,%
]{revtex4-2}
\usepackage{subfigure}
\usepackage{graphicx}
\usepackage{dcolumn}
\usepackage{bm}
\usepackage{comment}
\usepackage[utf8]{inputenc}
\usepackage[T1]{fontenc}
\usepackage{mathptmx}
\usepackage{etoolbox}

\makeatletter
\def\@email#1#2{%
 \endgroup
 \patchcmd{\titleblock@produce}
  {\frontmatter@RRAPformat}
  {\frontmatter@RRAPformat{\produce@RRAP{*#1\href{mailto:#2}{#2}}}\frontmatter@RRAPformat}
  {}{}
}%
\makeatother
\begin{document}

\preprint{AIP/123-QED}

\title{Analytical design of acoustic metasurface cells incorporating meander-line and Helmholtz resonators}
\author{Dingcheng Yang}
\author{Yan Kei Chiang}%
\author{David Powell}
 \email{dingcheng.yang@unsw.edu.au.}
\affiliation{ 
University of New South Wales, School of Engineering and Technology, Northcott Drive, Canberra,
ACT 2600, Australia.
}%

%


\begin{abstract}
Gradient acoustic metasurfaces have shown strong potential for manipulation of acoustic waves across the audible and ultrasonic frequency ranges. The key challenge in designing acoustic metasurfaces is to create a series of sub-wavelength unit cells that match the desired phase response. The most commonly used geometry is a series of Helmholtz resonators, side-coupled to a narrow channel. Despite the existence of a closed-form solution for 3 side-coupled resonators, most reported designs instead make use of 4 or more resonators, which require opaque, optimization-based design approaches. We show that the limiting factor in designs based on side-coupled resonators is the requirement to use elements with a positive imaginary part of impedance, which implies a Helmholtz resonator operating above its fundamental resonance - contradicting the requirement for sub-wavelength volume. We show that by replacing some of the Helmholtz resonators with meander-line elements, the required impedance values can readily be realized within a sub-wavelength volume. The metasurface design approach is demonstrated for a lens operating at 3 kHz and verified numerically. Furthermore, incorporating meander-line elements leads to improved broadband focusing performance, even when no explicit dispersion engineering is included in the design. As this design includes narrow channels, we include the effects of thermo-viscous losses in our modeling, and confirm that our design still gives superior performance to the reference design using only Helmholtz resonators. Lastly, the asymmetric case has been studied and the meander-line design shows a higher efficiency and excellent phase match, which is attributed to the better impedance matching. Our design is expected to lead to more optimal performance of acoustic metasurface designs, and the ability to make use of a closed-form design formula is expected to facilitate the analysis of fundamental performance bounds and enable more explicit achromatic design processes.

\end{abstract}

\maketitle

\section{Introduction}

Acoustic metasurfaces have been extensively investigated and developed over the past decades. Considerable research has been conducted into unit-cell structure designs to achieve wave manipulation functions. Many unit-cell topologies have been proposed and modified for phase tuning including Helmholtz resonators\cite{su_amplitude-modulated_2020,lu_achromatic_2023,guo_manipulating_2018}, labyrinthine cells \cite{qu_broadband_2022,zhang_acoustic_nodate,zou_reflected_2022} and membranes\cite{liu_magnetically_2020,sun_rabi-like_2024}. Thanks to artificial materials freely-tailorable shape and subwavelength size, they can realise many functions including beam steering\cite{li_arbitrarily_2020,song_frequency-selective_2021,chiang_reconfigurable_2020}, self-bending beams\cite{li_modulation_2021,tang_generation_2021}, acoustic lenses\cite{han_broadband_2021,wang_tunable_2020,peng_broadband_2022}, acoustic cloaking \cite{zhou_ultra-broadband_2021,zhou_tunable_2020}, sound attenuation and absorption\cite{ji_low-frequency_2020,guo_extremely-thin_2022,liu_dynamic_2021} and levitation\cite{stein_shaping_2022}.

The design of the unit-cells is often based on the Generalized Snell's law (GSL)\cite{li_reflected_2013}, which gives the required phase distribution along the metasurface for different wavefront manipulations. For designing the geometric parameters of the unit cells process\cite{li_modulation_2019,wang_tunable_2020,li_arbitrarily_2020}, a parametric sweep that has 2\(\pi\) span coverage is sufficient for the target phase profile. However, the design parameter space is not fully investigated and the relationship between the specific geometric parameter and phase change is ambiguous. Moreover, this kind of metasurface is usually narrow-band due to its limited degrees of freedom. Another design approach \cite{dong_achromatic_2022,zhou_ultra-broadband_2021}integrated with optimization algorithms with hundreds of degrees of freedom leads to another extreme and can indeed achieve a larger frequency band. But these unit-cells with peculiar shapes require high fabrication accuracy, making them even more sensitive to losses and less practical. Furthermore, such optimization-based design approaches give no insight into the physical constraints on performance.

If we consider the meta-atom design procedure from another perspective, impedance engineering\cite{monticone_full_2013}, the objective of constructing a metasurface is then converted to obtaining the corresponding impedance profile and finding appropriate geometry that meets the requirement. This approach also clearly demonstrates the phase and efficiency limits of the structure that be achieved if they can be converted to realizable impedance profiles. Research in electromagnetic metasurfaces shows that the three-layer impedance model explicitly finds the solutions for the impedance values and one can control the wave response well if the impedance of each layer is matched\cite{epstein_synthesis_2016}.  Li et al.~\cite{li_systematic_2018} proposed a bianisotropic acoustic metasurface consisting of four side-coupled Helmholtz resonators with different volumes and showed higher transmission efficiency compared to conventional symmetric designs. By characterizing the unit cell with an impedance matrix, their structure closely approximated the required two-port parameters and achieved enhancement in beam steering, especially at large angles. They also explained the necessity of using four Helmholtz resonators as geometric constraints prevent operation at or above resonance, which are needed to provide extreme impedance values.
Tong et al.~\cite{tong_asymmetric_2021} presented a unit-cell made from cascaded meander-lines of different geometries, fitting well with the required two-port parameters, and showing great improvement in transmission efficiency in large angle refraction. Therefore, impedance engineering for unit-cell design provides a different path that transfers the phase requirement into impedance fitting. It depicts a clearer relationship between the geometry and the phase and more degrees of freedom can better realise target wave response manipulation. However, it is still challenging to fit the impedance for three layers even knowing the exact value due to the geometry limitation.

In this work, we analyze the required impedance profile of a unit-cell via three-layer impedance model that aims for specific phase values with high transmitting efficiency. As we consider the design of lenses with low numerical aperture, and hence no large refraction angles, our lens is based on a symmetric design of the unit cell. An inverse design process is applied to identify the target impedance of each layer. We propose a symmetric unit-cell with the meander-line resonator in the center and identical Helmholtz resonators on two sides. The impedance libraries of the  Helmholtz resonator and meander-line resonator are established, considering restrictions on available space. It is shown that the meander-line cell can realize a much broader impedance coverage than the Helmholtz resonator under the same space constraint. We design the metalens and compare our design to the three Helmholtz side-coupled structure and find a higher acoustic intensity enhancement (6.0 versus 3.5) at the focal spot over a large frequency range in numerical simulations. When including the effects of thermo-viscous losses, we show that our design still achieves superior performance to the Helmholtz resonator designs, despite having higher absorption losses. To confirm the generality of our method, we design an asymmetric unit cell exhibiting acoustic bianisotropy (also known as Willis coupling), and show that the the meander-line structure has superior performance in this case too.

\section{\label{sec:level2}Unit cell architecture}

\begin{figure*}[hbt]
    \centering
    \subfigure{\includegraphics[width=0.38\textwidth,]{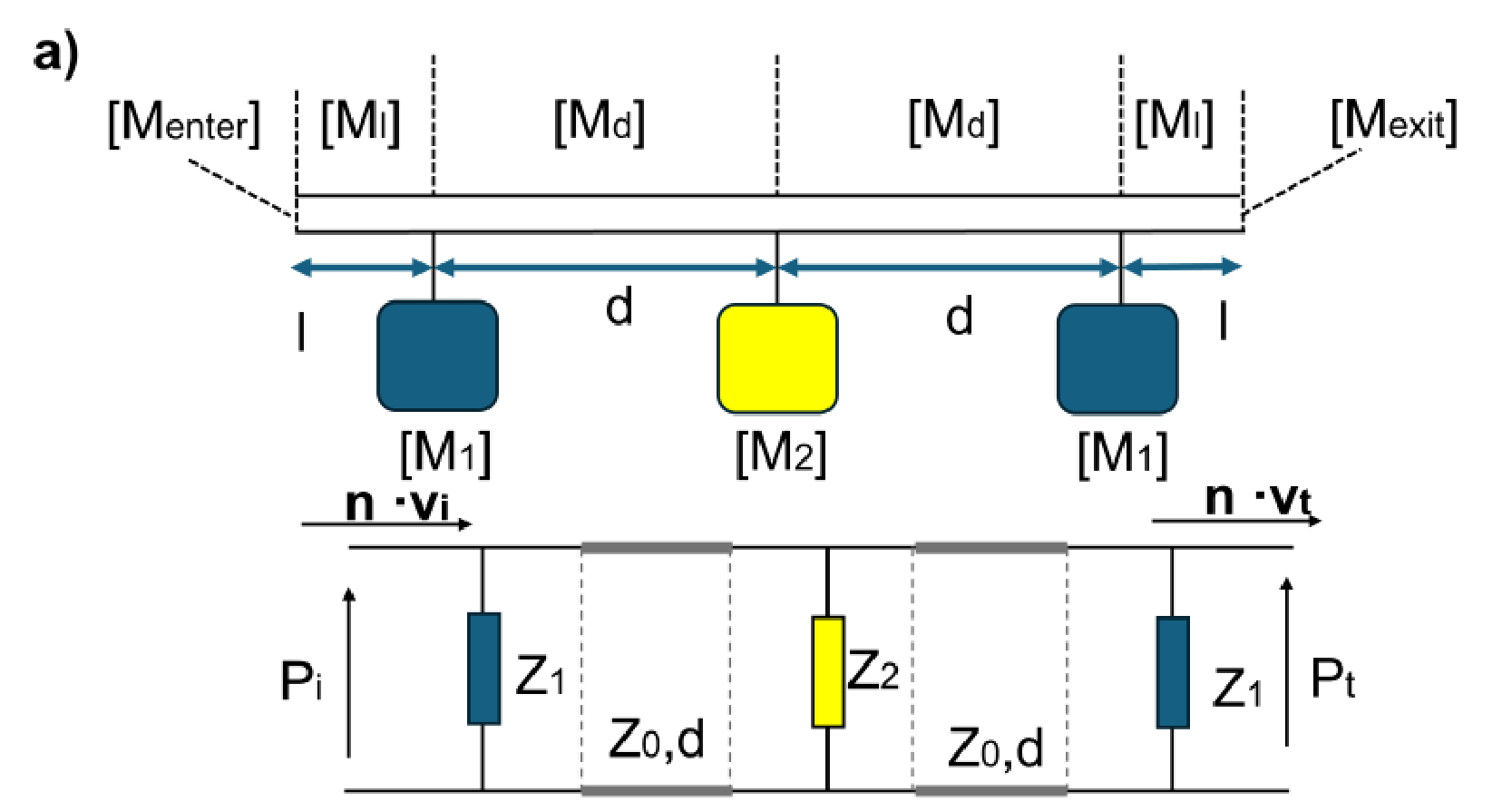}}
    \subfigure{\includegraphics[width=0.32\textwidth,]{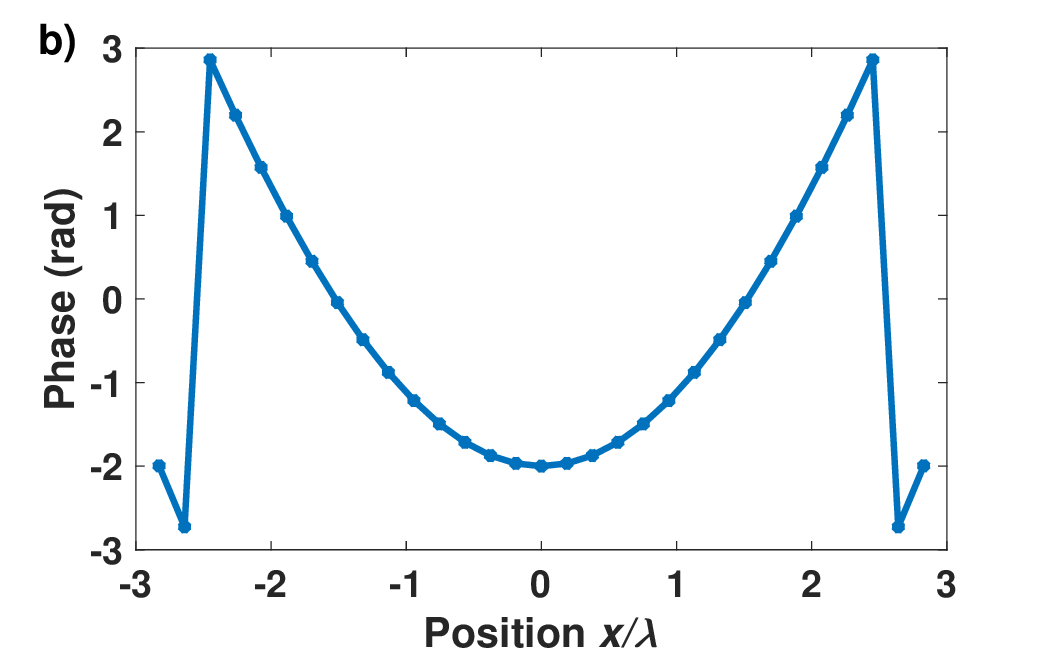}}

\caption{(a) The diagram of the three side coupled impedance unit-cell with equal spacing $d$ and same distance to the end $l$ and the transfer matrix of each part. The bottom shows its equivalent circuit. In our case, impedance in first and third positions is equal due to symmetry (b) The target phase profile over the 31-unit metalens.}
  \label{fig:1}
\end{figure*}

The geometry of the unit cell is shown in Fig.~\ref{fig:1}a, consisting of a narrow channel with 3 side-coupled resonators. We apply the transfer matrix method to model this structure, and to find the required impedance of the 3 resonators. For the incoming wave and refracted wave through the metasurface, their pressure fields can be expressed as:
\begin{eqnarray}\label{incident}
p_i(x,y)=p_0e^{-ik(\sin\theta_i x+\cos\theta_i y)};
\end{eqnarray}
\begin{eqnarray}\label{refraction}
p_t(x,y)=Tp_0e^{-ik(\sin\theta_t x+\cos\theta_t y)};
\end{eqnarray}

where \(p_0\) is the amplitude of the incident wave, \(\theta_i\) and \(\theta_t\) are the incident and transmitted angles, respectively. \(k\) is the wave number in background medium and \(T\) is the transmission coefficient. The velocity vector of two sides of the metasurface can be expressed as:

\begin{eqnarray}\label{incident velocity}
\vec{v_i}(x,y)=\frac{p_i(x,y)}{Z_0}(\sin\theta_i \vec{x}+\cos\theta_i \vec{y});
\end{eqnarray}
\begin{eqnarray}\label{refraction velocity}
\vec{v_t}(x,y)=\frac{p_t(x,y)}{Z_0}(\sin\theta_t \vec{x}+\cos\theta_t \vec{y});
\end{eqnarray}

where \(Z_0=\rho_0 c_0\) is the characteristic impedance of the background medium. To investigate the local phase response on the metasurface, the unit cell is modelled via transmission line theory\cite{diaz-rubio_acoustic_2017} and the two-port transfer matrix \([M]\) for each unit cell can be expressed as:

\begin{eqnarray}\label{ABCD matrix}
\begin{bmatrix}
p_i(x,0) \\[5pt] \hat{n}\cdot\vec{v_i}(x,0)
\end{bmatrix}
=
\begin{bmatrix}
M_{11} & M_{12}\\[5pt]
M_{21} & M_{22}
\end{bmatrix}
\begin{bmatrix}
p_t(x,H) \\[5pt] \hat{n}\cdot\vec{v_t}(x,H)
\end{bmatrix}
\end{eqnarray}

where \(\hat{n}\) is the normal vector of the metasurface and \(H\) is the total height of the metasurface.

As shown in Fig.~\ref{fig:1}(b), our proposed symmetric unit cell can be realized by a pipe having three side-coupled resonators separated by distance $d$ and $l$ is the distance from the resonator to the end. We can obtain the total transfer matrix $[M_t]$ by cascading all elements in the unit-cell as:

\begin{eqnarray}\label{total M}
    [M_t]=[M_{enter}][M_{l}][M_1][M_{d}][M_2][M_{d}][M_1][M_{l}][M_{exit}]
\end{eqnarray}

where $[M_{enter}]$ and $[M_{exit}]$ are transfer matrices accounting for the change of channel width, $[M_{l}]$ is the transfer matrix of the transmission line of $l$ and $[M_{d}]$ is the transmission line between each resonator and $[M_{1}]$, $[M_{2}]$ are the transfer matrices of first and second resonators, as the third one is identical to the first one due to symmetry.

For a symmetric unit cell, the transmission and reflection coefficients of forward (\(t^+, r^+\)) and backward incident (\(t^-, r^-\)) waves are the same. The scattering matrix $[S]$ consists of these coefficients and can be determined if we set the target phase value, unity amplitude for transmission and zero reflection. The $[S]$ matrix can be converted to a total transfer matrix $[M_t]$ by:

\begin{eqnarray}\label{S matrix}
\begin{bmatrix}
S_{11} & S_{12}\\[5pt]
S_{21} & S_{22}
\end{bmatrix}
=
\begin{bmatrix}
r^+ & t^+\\[5pt]
r^- & t^-
\end{bmatrix}
\end{eqnarray}

\begin{eqnarray}\label{s to abcd}
{\begin{bmatrix}

M_{11} & M_{12}\\[5pt]
M_{21} & M_{22}
\end{bmatrix}_t}
=
\begin{bmatrix}
\frac{(1+S_{11})(1-S_{22})+S_{12}S_{21})}{2S_{21}} & Z_0\frac{(1+S_{11})(1+S_{22})-S_{12}S_{21})}{2S_{21}}\\[5pt]
\frac{1}{Z_0}\frac{(1-S_{11})(1-S_{22})-S_{12}S_{21})}{2S_{21}} & \frac{(1-S_{11})(1+S_{22})+S_{12}S_{21})}{2S_{21}}
\end{bmatrix}
\end{eqnarray}
where $Z_0$ is the impedance of the background.

The use of 3 resonators allows a closed-form solution, where the impedance $Z_n$ of each layer can be directly related to the two-port impedance parameters. This solution can be applied if a transfer matrix $[M_{sub}]$ starting exactly from the first resonator and with no width change which is part of $[M_t]$. And it can be expressed as:

\begin{eqnarray}\label{sub M}
    [M_{sub}]=[M_{enter}]^{-1}[M_{l}]^{-1}[M_{t}][M_{l}]^{-1}[M_{exit}]^{-1}
\end{eqnarray}

Then, we need to convert $[M_{sub}]$ to impedance matrix [Z] as:
\begin{eqnarray}\label{ABCD to Z matrix}
\begin{bmatrix}
Z_{11} & Z_{12}\\[5pt]
Z_{21} & Z_{22}
\end{bmatrix}
=
{\begin{bmatrix}
\frac{M_{11}}{M_{21}} & \frac{M_{11}M_{12}-M_{12}M_{21}}{M_{21}}\\[5pt]
\frac{1}{M_{21}} & \frac{M_{22}}{M_{21}}
\end{bmatrix}_{sub}}
\end{eqnarray}

And we can finally obtain the required impedance from the closed-form solution as\cite{wong_reflectionless_2016}:

\begin{eqnarray}
Z_1 &=&\frac{Z_r\det[Z]\sin(kd)}{i|Z|\cos(kd)+Z_r(Z_{12}+Z_{22})\sin(kd)} \label{3_layer_Z1}\\
Z_2 &=&\frac{Z_r^2Z_{12}\cos(2kd-1)}{-i2Z_rZ_{12}\sin(2kd)+2|Z|} \label{3_layer_Z2}\\
Z_3 &=&\frac{Z_r\det[Z]\sin(kd)}{i|Z|\cos(kd)+Z_r(Z_{12}+Z_{11})\sin(kd)} \label{3_layer_Z3}  
\end{eqnarray}
where \(Z_r\) is the impedance of the narrow pipe, \(i\) is the unit imaginary number, \(|Z|\) is the determinant of the impedance matrix and $k=\omega/c$ is the wavenumber in the pipe connecting the resonators. Noting that Eqs.~\eqref{3_layer_Z1} and \eqref{3_layer_Z3} differ only in the appearance of \(Z_{11}\) or \(Z_{22}\) in the denominators. For this design we consider a symmetric cell with \(Z_{11}=Z_{22}\), hence the required shunt impedances are identical, \(Z_1=Z_3\). In Section \ref{sec:asymmetric} we show that our design is also application for an asymmetric unit cell with $Z_1\neq Z_3$.

\section{\label{sec:results}Metasurface Lens}

\subsection{Lens Design}

     \begin{figure*}[ht]
    \centering
    \subfigure{\includegraphics[width=0.2\textwidth]{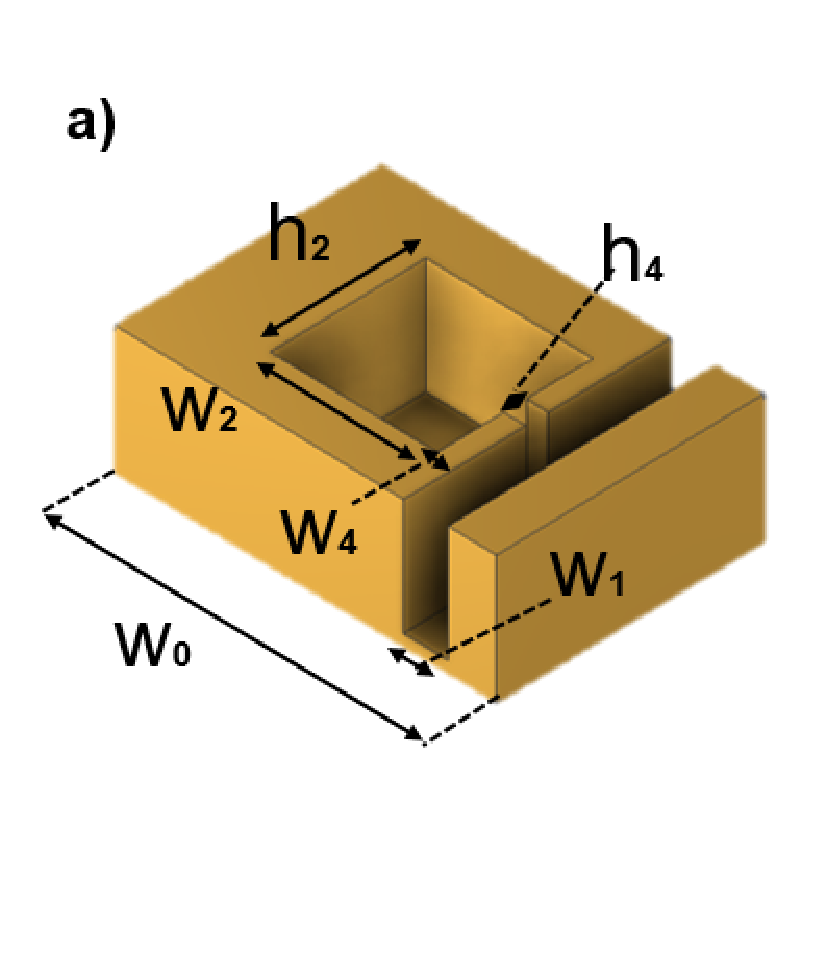}} 
    \subfigure{\includegraphics[width=0.29\textwidth]{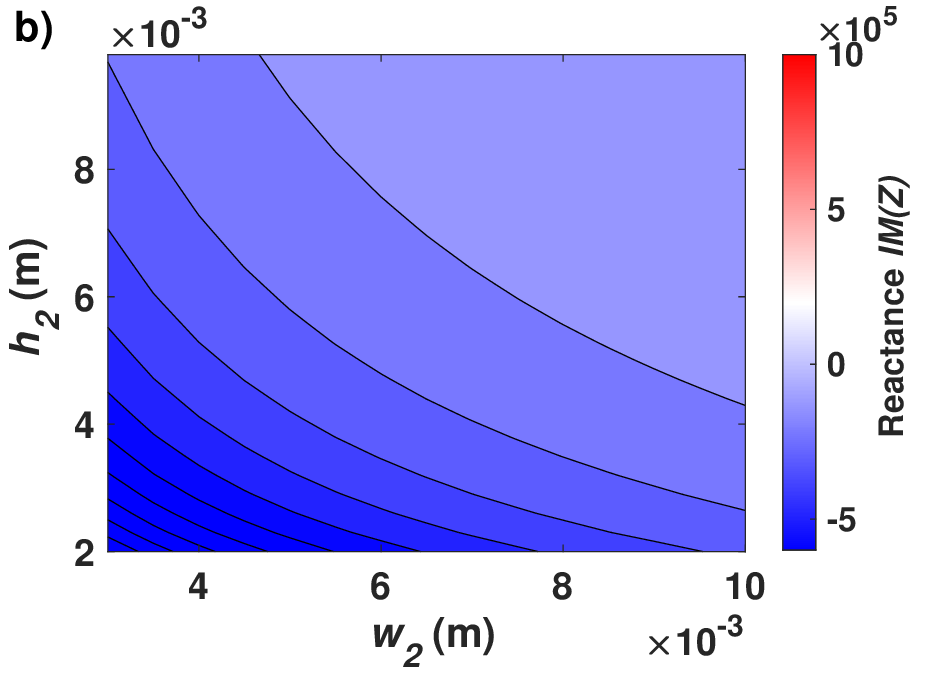}}
    \subfigure{\includegraphics[width=0.2\textwidth]{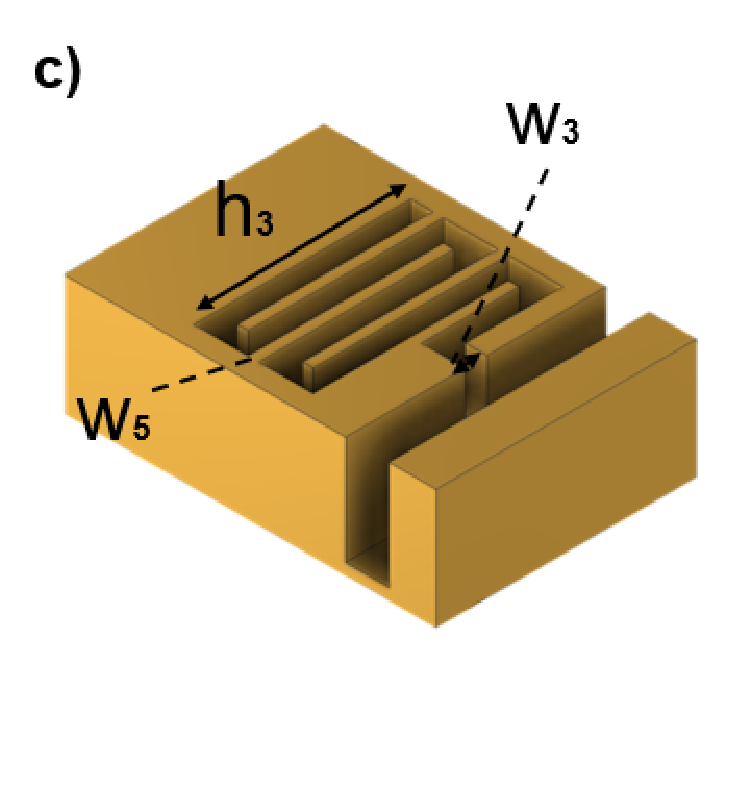}}
    \subfigure{\includegraphics[width=0.29\textwidth]{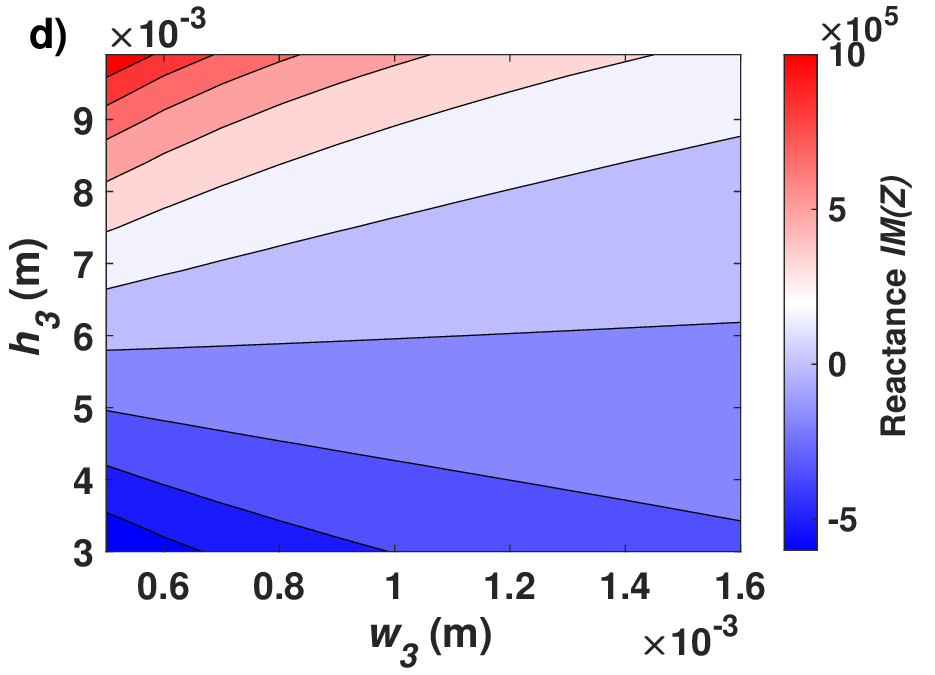}}
\caption{(a) The Helmholtz resonator. (b)  The simulated imaginary part of the impedance (Reactance) library of the Helmholtz resonator. (c) The meander-line resonator. (d)  The simulated imaginary part of the impedance (Reactance) library of the meander-line resonator.  }
\label{fig:2}
\end{figure*}
\begin{figure*}[ht!]
    \centering
    \subfigure{\includegraphics[width=0.4\textwidth]{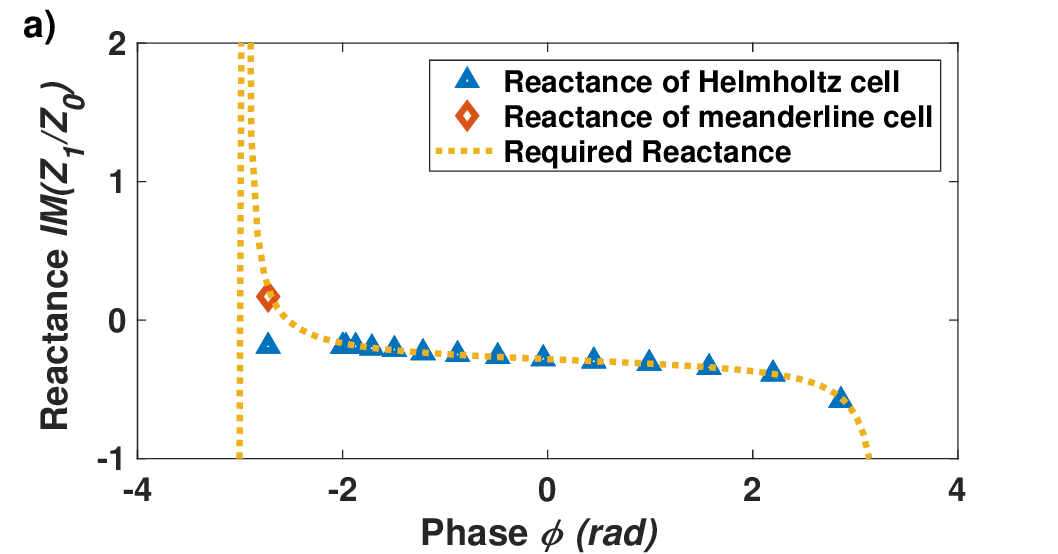}}
    \subfigure{\includegraphics[width=0.4\textwidth]{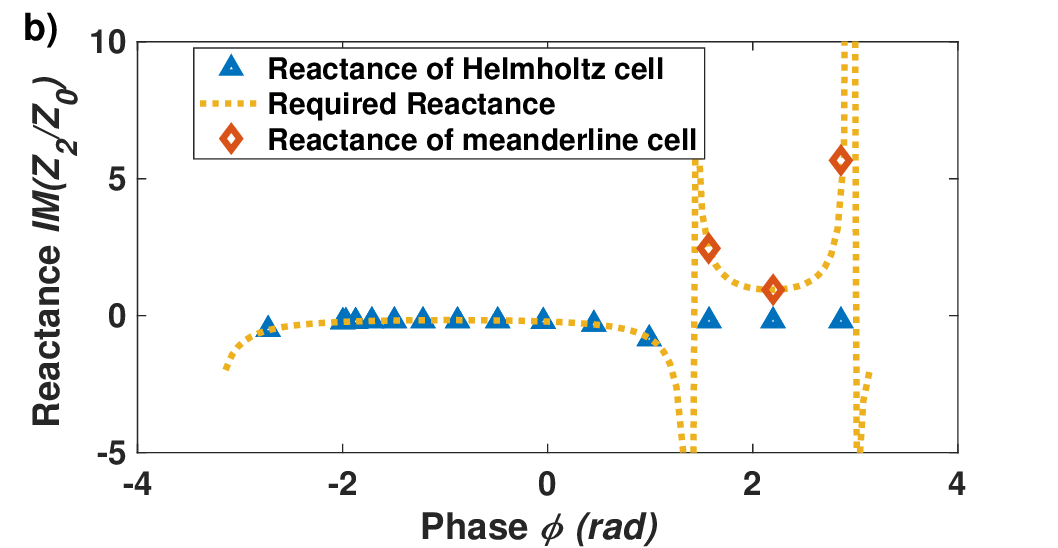}}
\caption{ The required impedance (dash line) over phase and the impedance fitted by Helmholtz cells(triangles) and by meander-line cells(diamonds) for \(Z_1\)(a) and \(Z_2\)(b), \(Z_0=2.0663e5\) is the impedance of the narrow pipe.}
\label{fig:3}
\end{figure*}

According the Generalized Snell's law, the phase profile for a metalens follows a hyperbolic distributions\cite{assouar_acoustic_2018}:
\begin{eqnarray}\label{GSL}
    \phi(x,\omega)=\frac{\omega}{c}(\sqrt{x^2+F^2}-F);
\end{eqnarray}
where \(\omega\) is the angular frequency, \(F\) is the focal length and \(c\) is the speed of sound. 
Fig.~\ref{fig:1}b shows the target phase profile for a 31-cell metalens with a focal length $F=400 mm$. Each unit cell is realised as three side-coupled resonances, as shown in Fig.~\ref{fig:1}b, with each resonator designed to match the three shunt impedances given by Eqs.~\eqref{3_layer_Z1}-\eqref{3_layer_Z3}.  
\begin{figure*}[t]
    \centering
    \subfigure{\includegraphics[height=7.0cm]{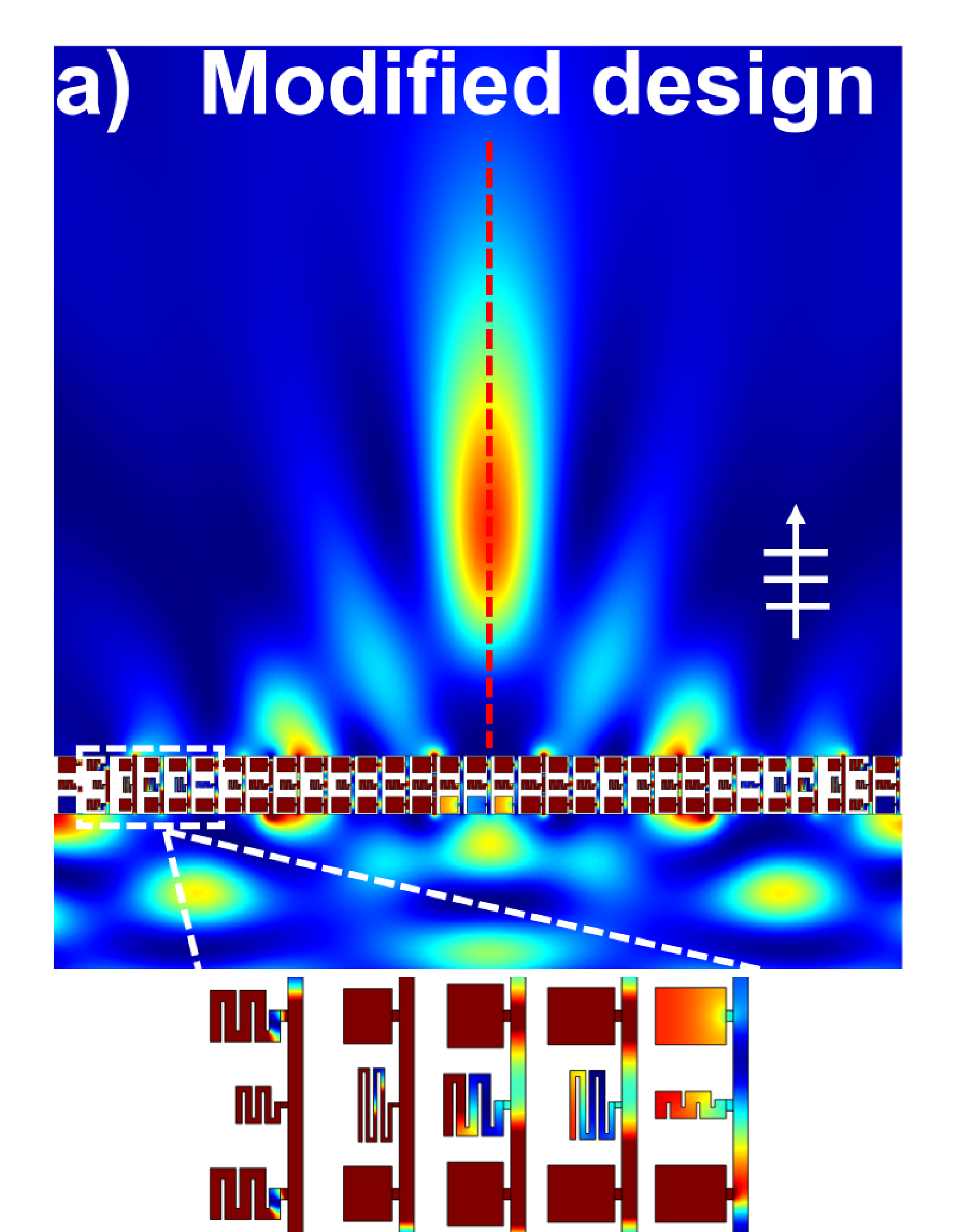}}
    \subfigure{\includegraphics[height=7cm]{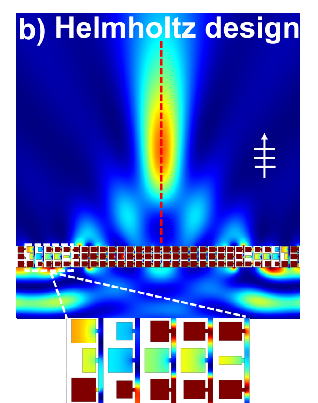}}
    \subfigure{\includegraphics[height=7cm]{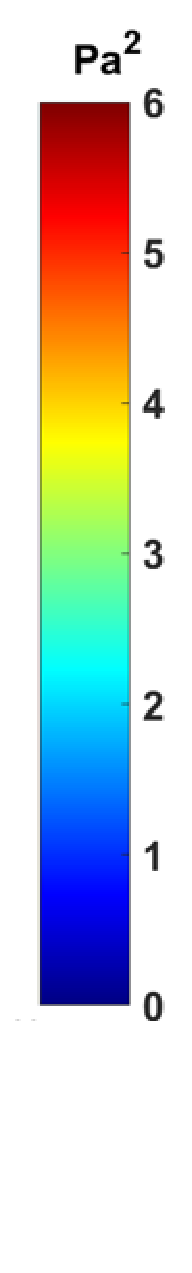}}
    
\caption{Top panels: The transmission intensity field of the metalens with (a) modified design at 2990 Hz (b) Helmholtz design at 3000 Hz. Bottom panels: the geometry of the 2nd to 6th unit-cell of the metasurface with (a) modified design and (b) Helmholtz design.}
\label{fig:4}
 \end{figure*}

\begin{figure*}[t]
    \centering
    \subfigure{\includegraphics[width=0.48\textwidth]{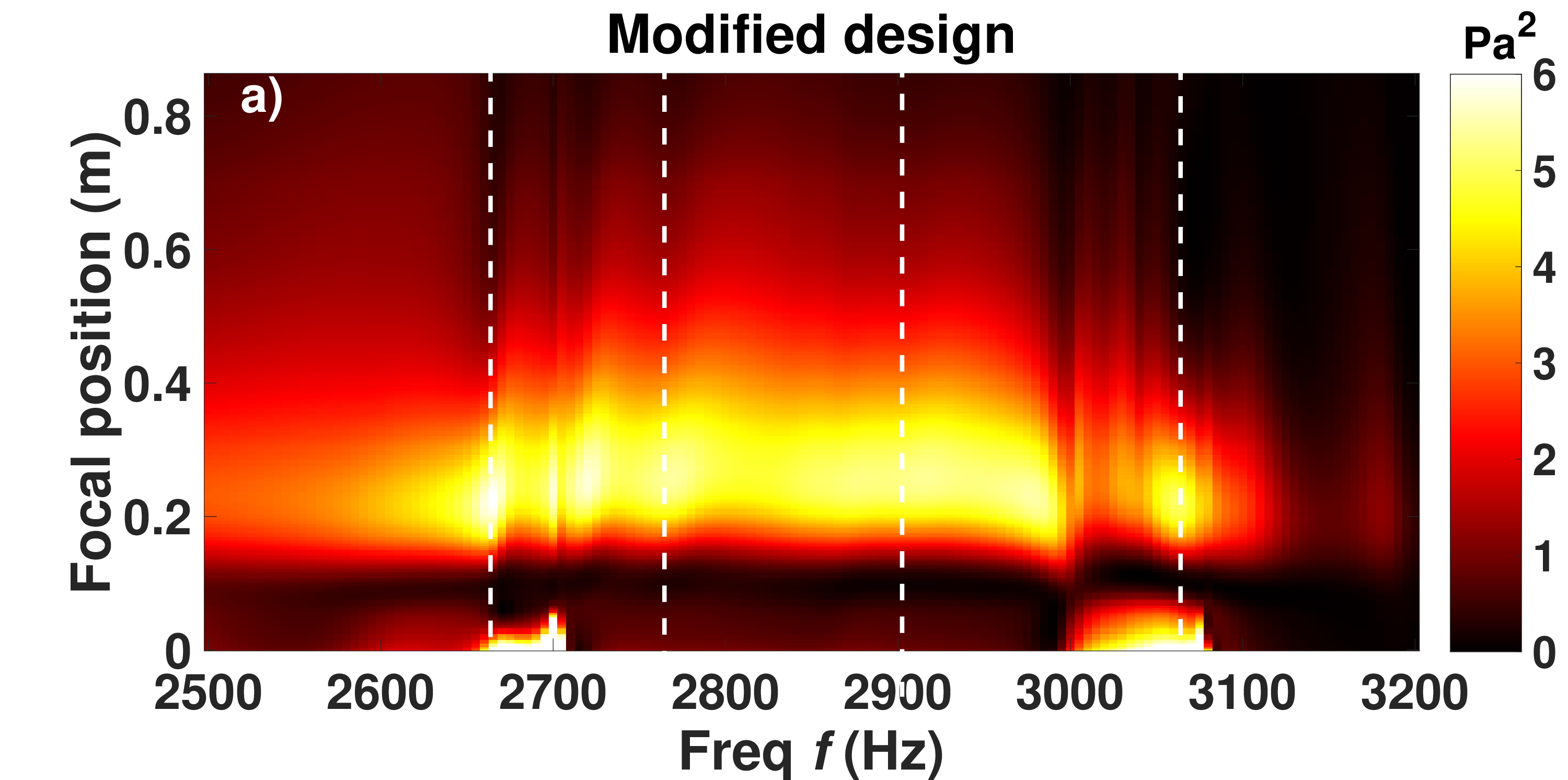}}
    \subfigure{\includegraphics[width=0.48\textwidth]{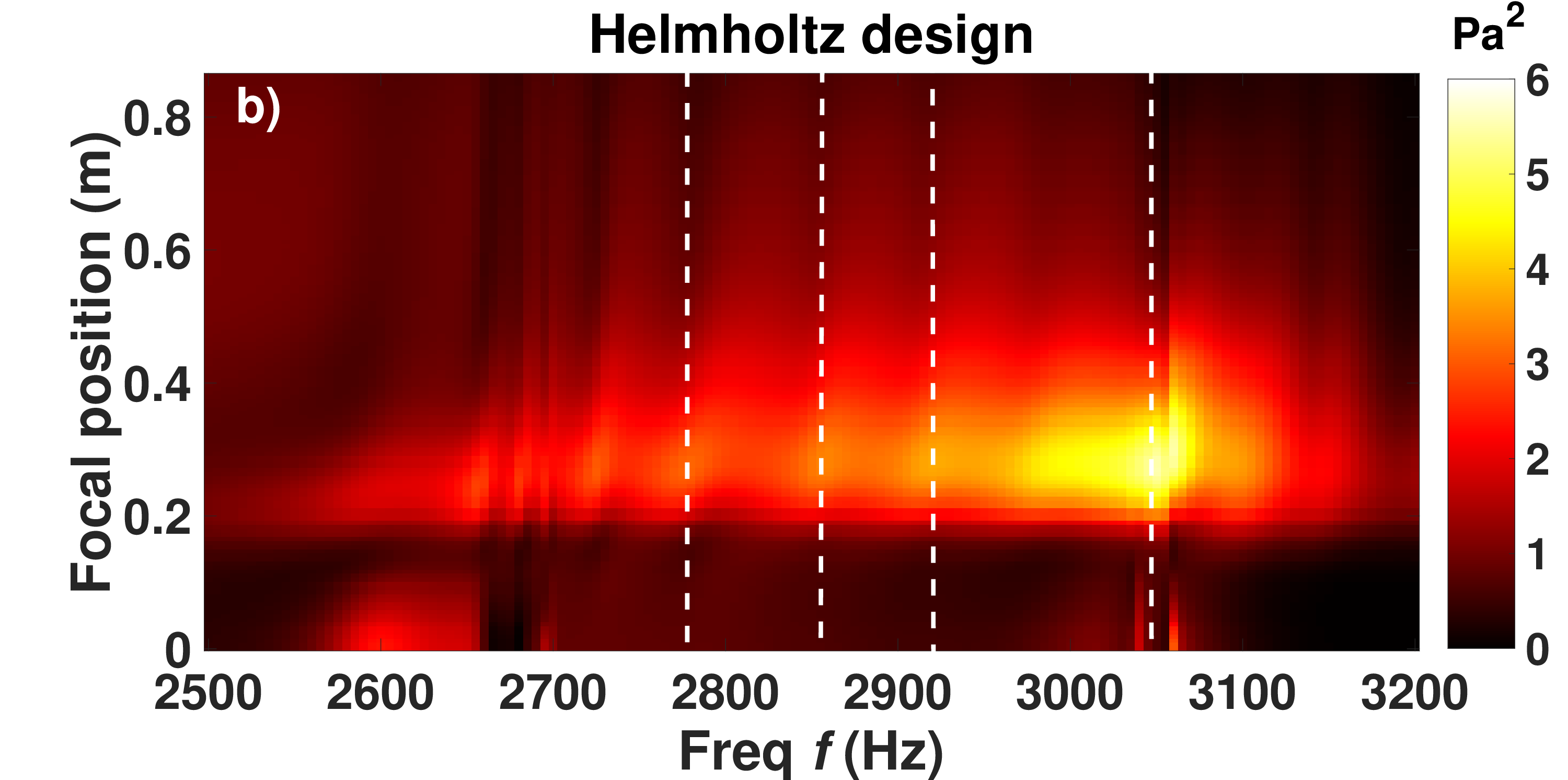}}
    \subfigure{\includegraphics[width=0.12\textwidth,height=2.65cm]{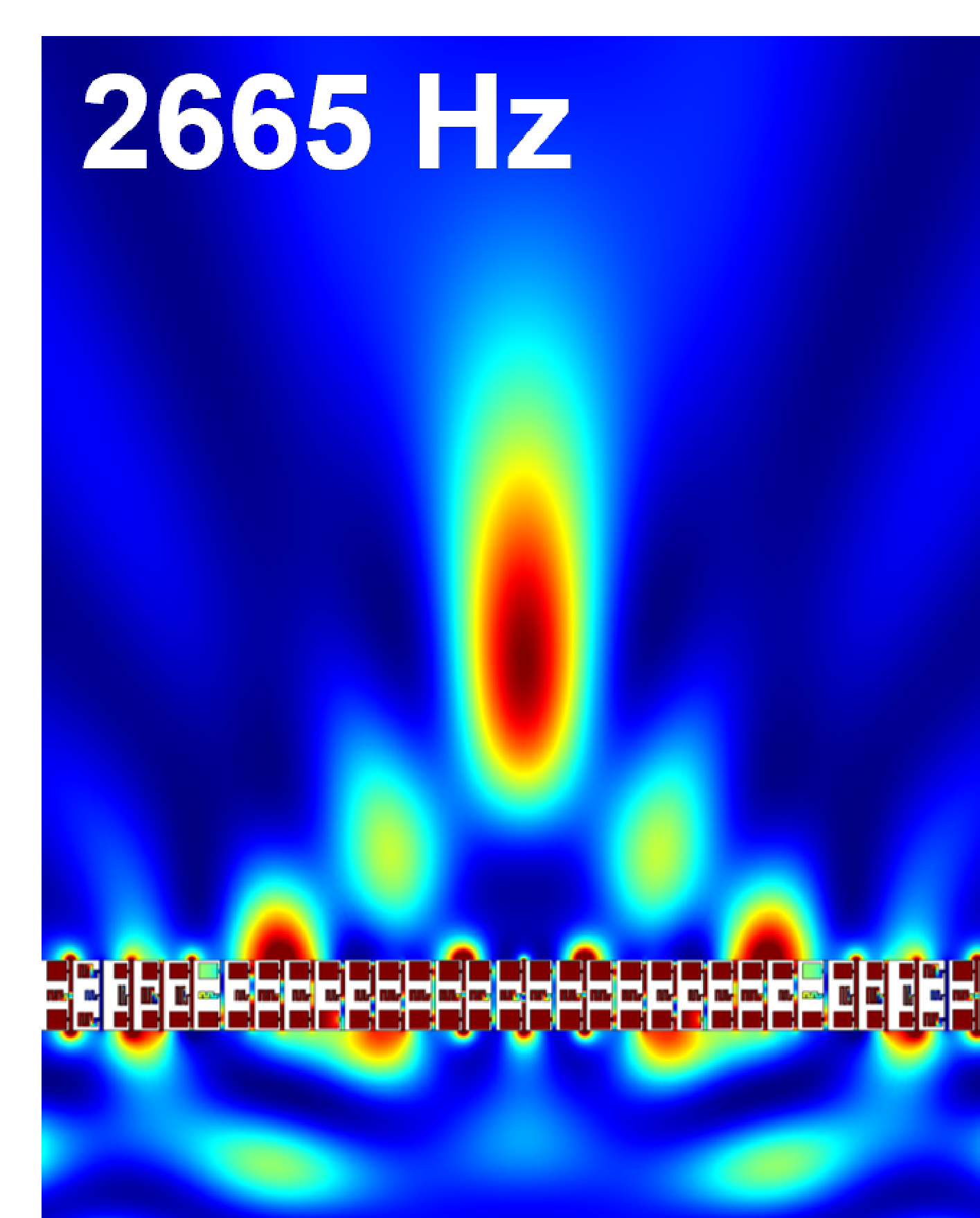}}
    \subfigure{\includegraphics[width=0.12\textwidth,height=2.65cm]{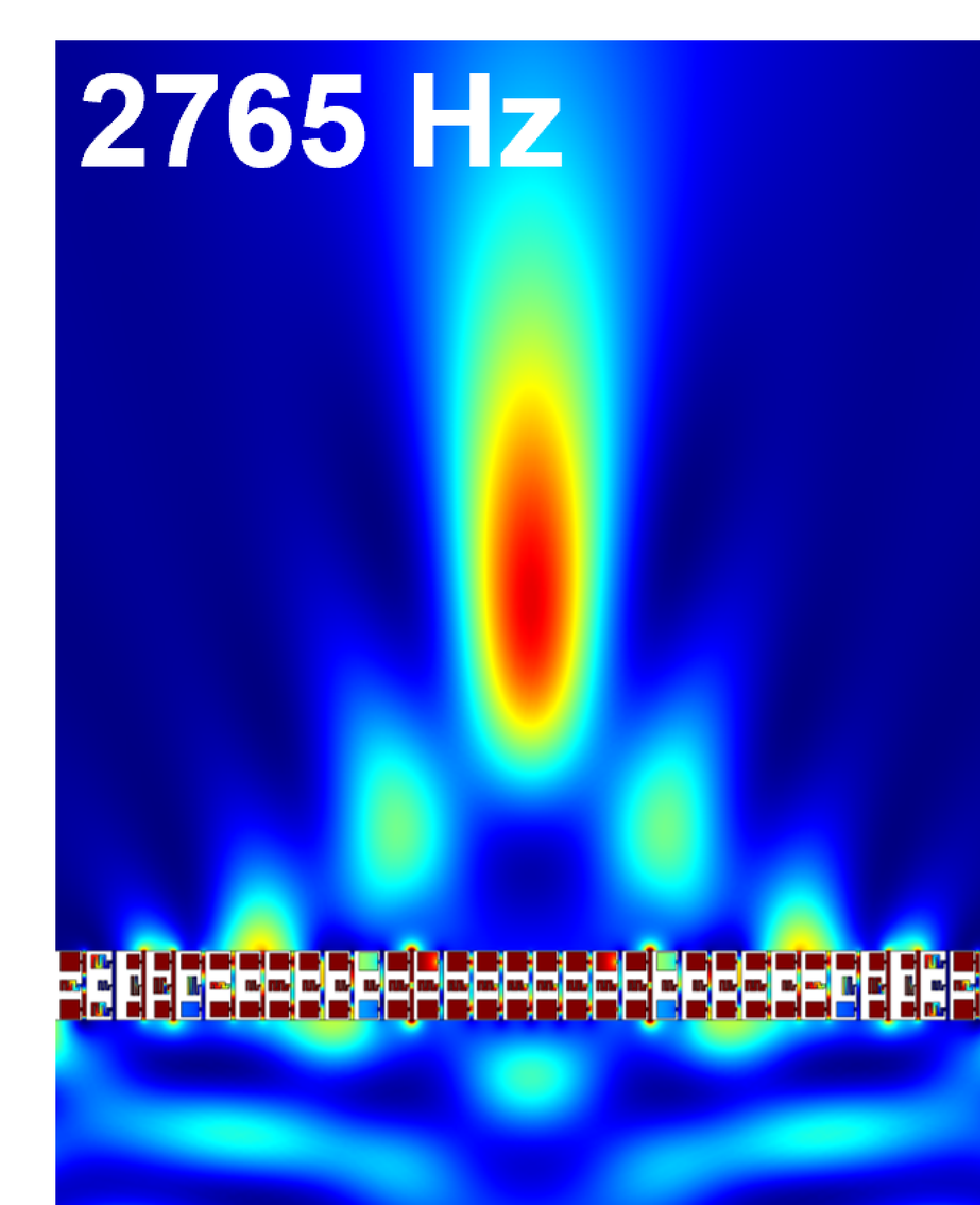}}
    \subfigure{\includegraphics[width=0.12\textwidth,height=2.65cm]{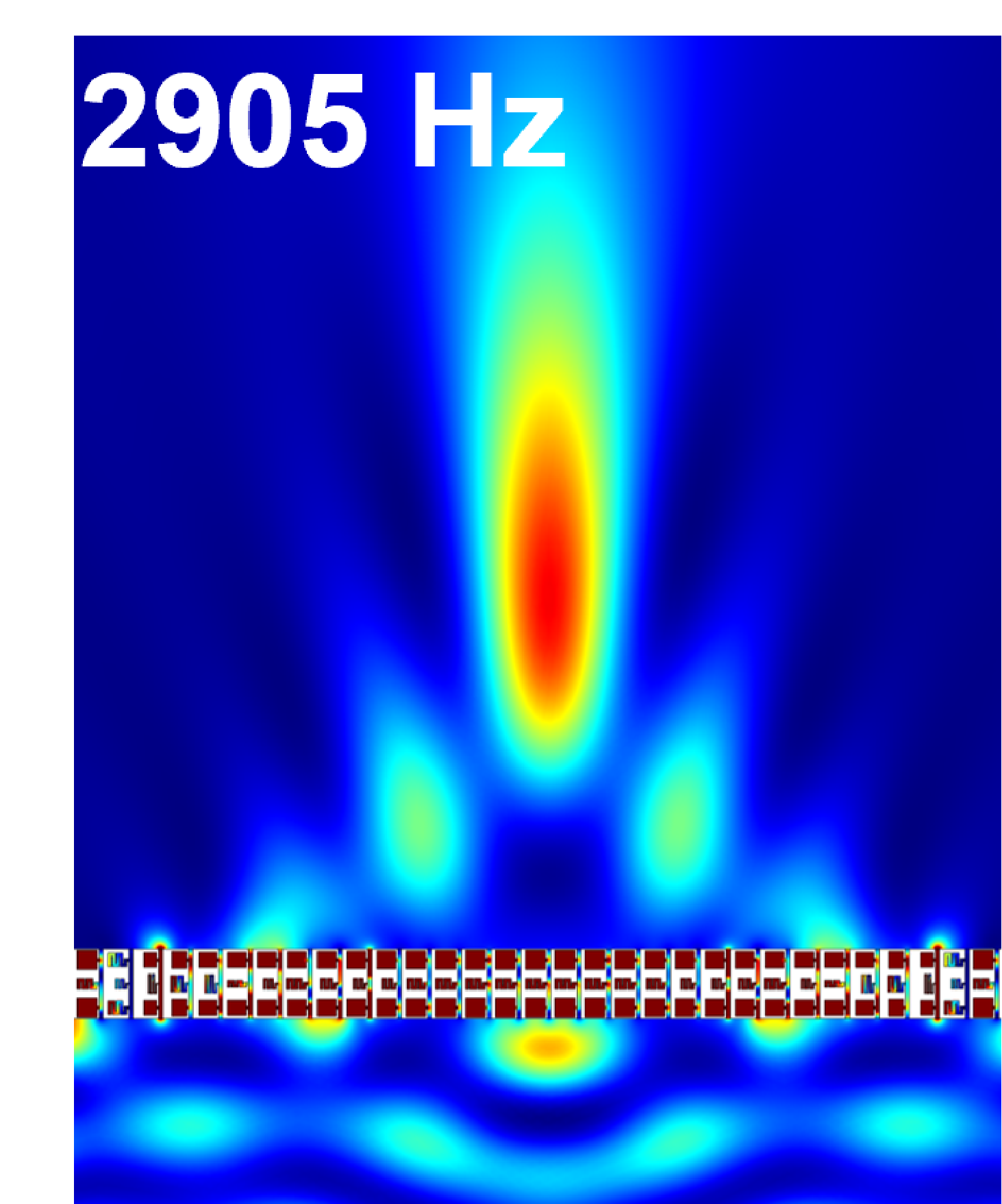}}
    \subfigure{\includegraphics[width=0.12\textwidth,height=2.65cm]{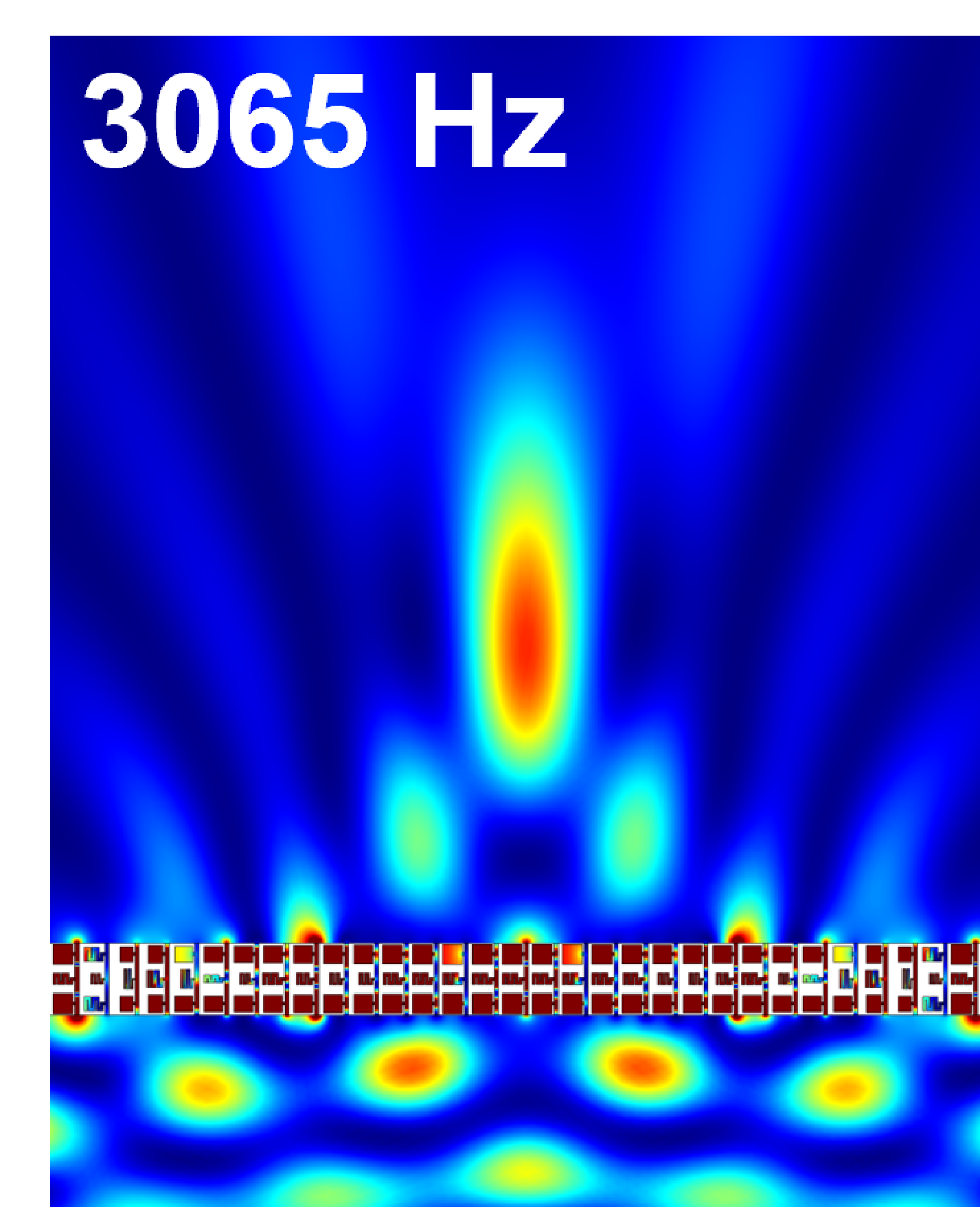}}
    \subfigure{\includegraphics[width=0.12\textwidth,height=2.61cm]{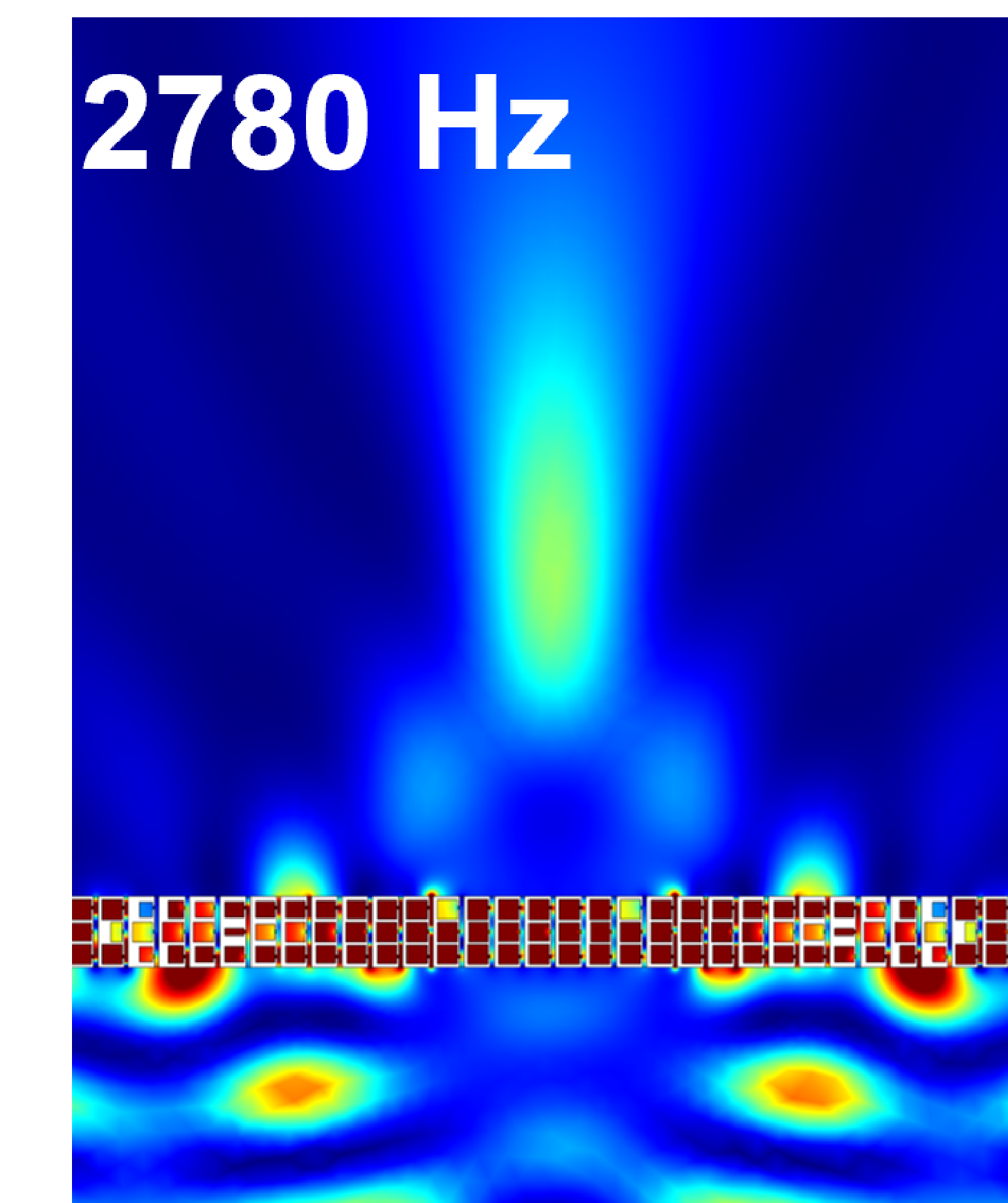}}
    \subfigure{\includegraphics[width=0.12\textwidth,height=2.61cm]{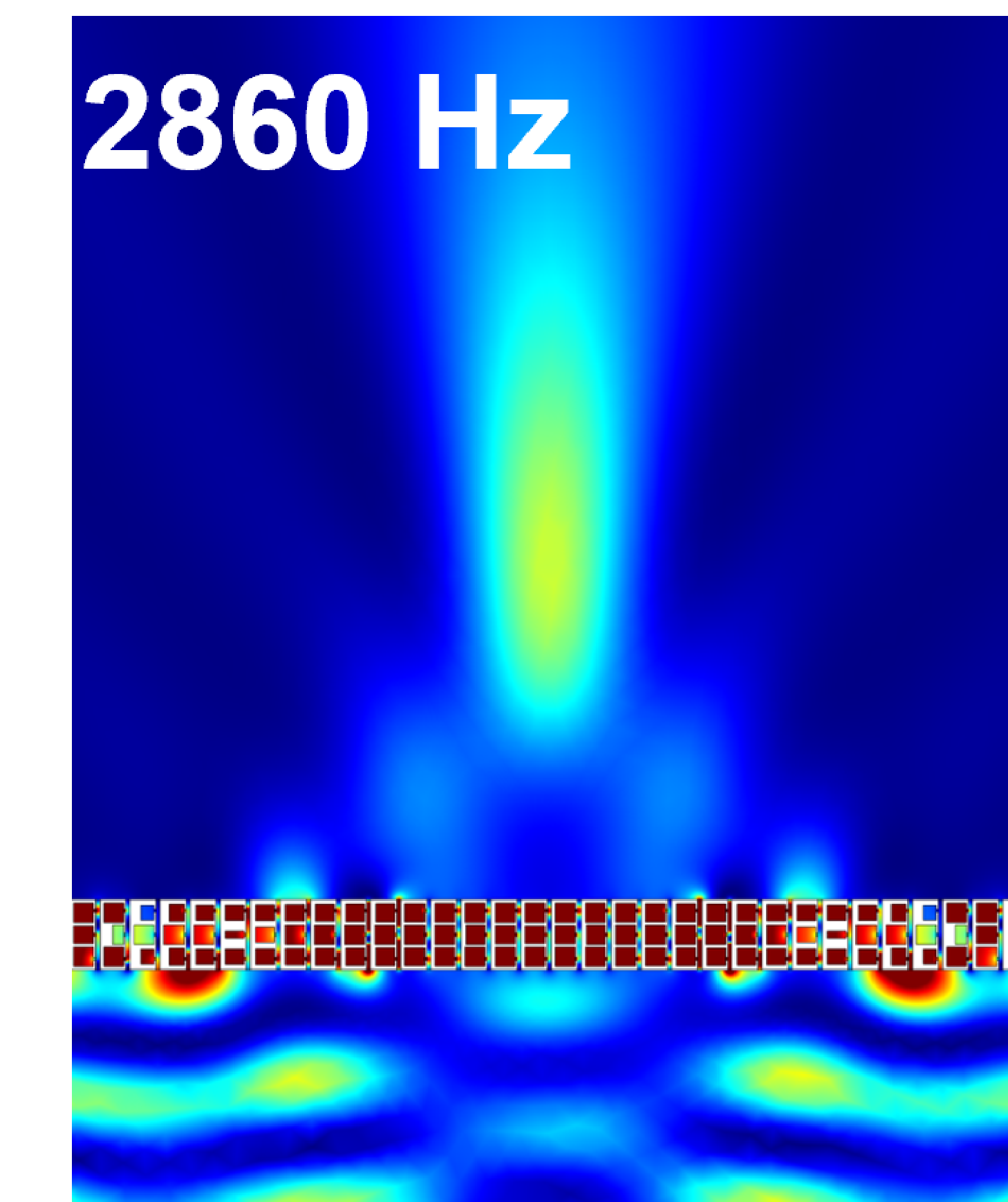}}
    \subfigure{\includegraphics[width=0.12\textwidth,height=2.61cm]{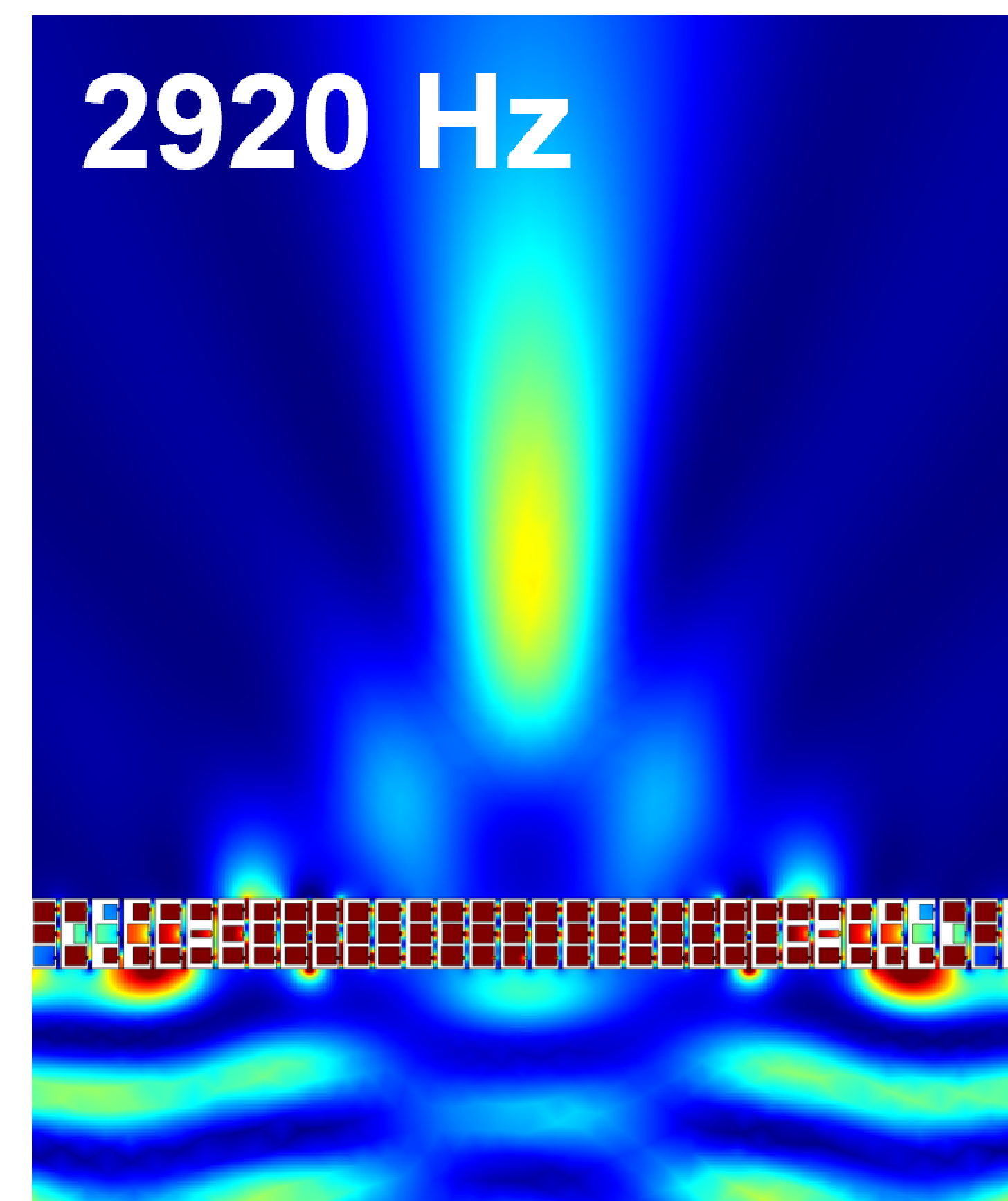}}
    \subfigure{\includegraphics[width=0.12\textwidth,height=2.61cm]{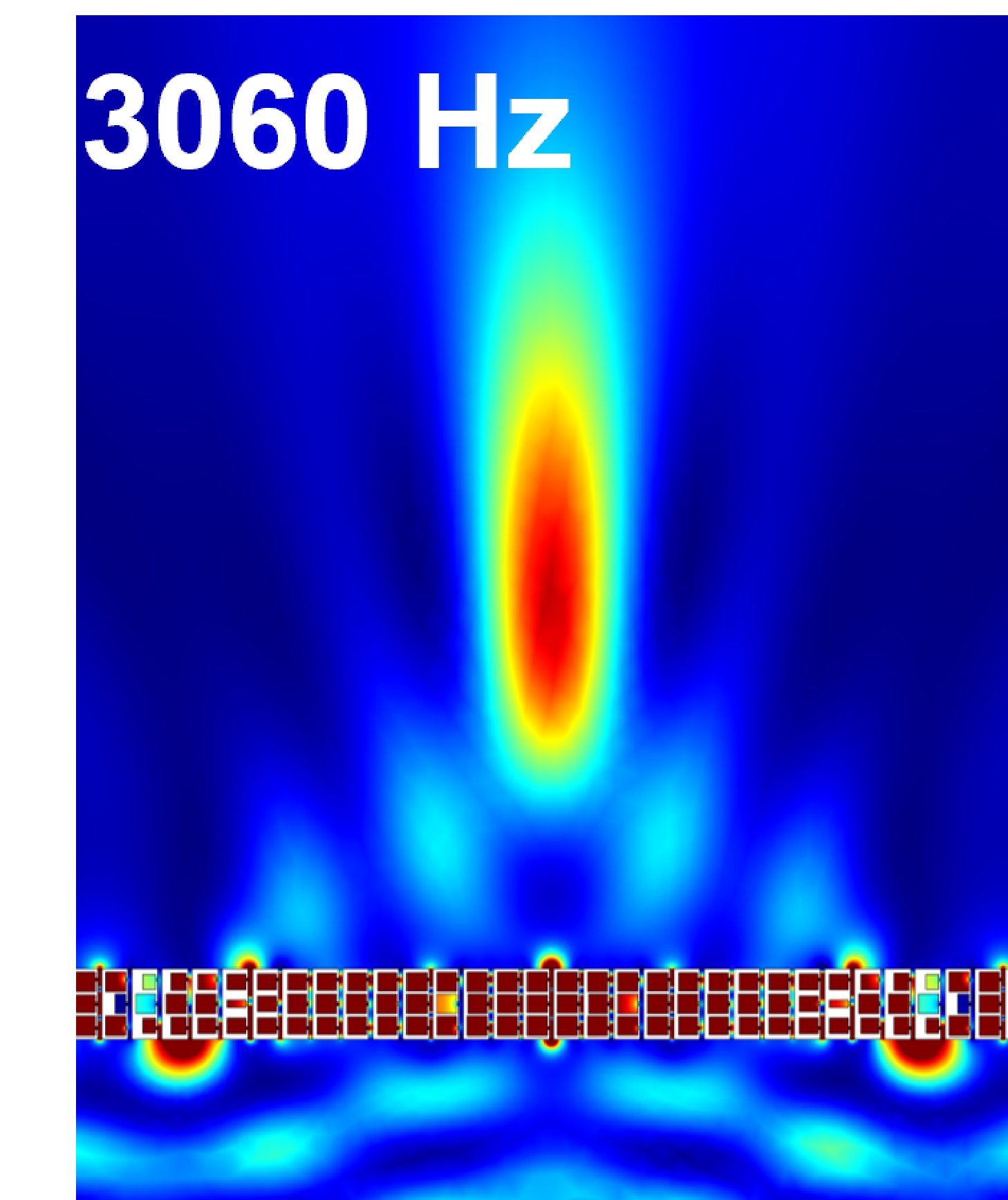}}

\caption{(a) Top: The power intensity coefficient plot of the central line of the modified metalens (shown as red-dot lines in Fig.~\ref{fig:4}) in the frequency range from 2.5 kHz to 3.2 kHz; white dash lines indicate four frequencies achieving high pressure enhancement. Bottom: The power intensity field of the modified metalens at 2660 Hz, 2730 Hz, 2860 Hz and 3060 Hz, respectively. (b) Top: The power intensity coefficient plot of the central line of the Helmholtz metalens in the frequency range from 2.5 kHz to 3.2 kHz; white dash lines indicate four frequencies achieving high pressure enhancement. Bottom: The power intensity field of the modified metalens at 2760 Hz, 2860 Hz, 2920 Hz and 3060 Hz, respectively.}
\label{fig:5}
 \end{figure*}
 
First we build the unit-cell using Helmholtz resonators, each having geometric parameters illustrated in Fig.~\ref{fig:2}a. We set the space constraint of the cavity for $10 mm$ for a deep-subwavelength unit-cell design. Here \(w_0=15mm\)  is the width of the unit cell, \(w_1=2 mm\) is the width of the side coupling air duct, \(w_4= 1 mm \) is the length of the neck of the resonator, \(h_4= 1.5 mm\) is the width of the neck, \(w_2\) the width of the cavity and \(h_2\) the height of the cavity are two variables which will be parameterized to obtain the required impedances. The 2D lossless simulations are conducted in finite element software \textit{COMSOL Multiphysics} at 3.0 kHz and the transmission and reflection coefficients are collected. They are converted to impedance and both real and imaginary parts can be found due to numerical errors. The real part is neglected as it is two orders of magnitude smaller than the imaginary part. The contour of the imaginary part of impedance (reactance) library of the Helmholtz resonator element is shown in Fig.~\ref{fig:2}b. A clear observation is that the reactance stays negative in this parameter space, because the sub-wavelength Helmholtz resonator is below its fundamental resonant frequency (4.4 kHz for the largest cavity). We attempt to fit the reactance to the required impedance profile shown in Fig.~\ref{fig:3}a,b. In Fig.~\ref{fig:3}a, most of the required transmission phase values can be well-approximated by the Helmholtz resonator-based design, except for one phase value near -2.8 rad. However, when positive reactance is required in \(Z_2\), three phase values cannot be achieved near 2 rad. As this figure only shows half of the elements in the metalens due to symmetry, the number of mismatching elements is doubled in the full metasurface.

To better approximate the full range of required impedance values in \(Z_1\) and \(Z_2\), we proposed the meander-line resonator. The structure is illustrated in Fig.~\ref{fig:2}c. $w_5=0.5 mm$ is the width of the teeth, \(w_3\) is the width of the meander-line duct and \(h_3\) is the height of the cell which has the same maximum value as \(h_2\) of the Helmholtz resonator. All other parameters \(w_0,w_1,w_4\) are the same as the Helmholtz resonator. The meander-line duct starts from exactly the neck of the Helmholtz resonator and winds four and a half times. \(w_3\) has a maximum of $1.6 mm$ to ensure it does not exceed the width \(w_2\) of the Helmholtz resonator. Fig.~\ref{fig:2}d shows the contour of the reactance library of the meander-line resonator. The coverage of reactance is almost doubled and positive values can be obtained, since the duct can be designed to operate above its fundamental resonance, which is 1.8 kHz for the largest structure that fits within our geometric constraints. Therefore, we apply three meander-line cells only in the cells of -2.725 rad and also substitute all the central elements with meander-line cells because the Helmholtz resonator cannot provide the required impedance. Fig.~\ref{fig:3}a and Fig.~\ref{fig:3}b show that the required impedance values are well-matched for all unit cells, and that complete coverage of required transmission phase values is achieved.

To compare the effect of the improved unit cell design on overall measurface performance, we design two meta-lenses, using the two types of cells shown in Fig.~\ref{fig:2}c and Fig.~\ref{fig:2}a. The first and third impedance elements of both cells are identical Helmholtz resonators. The modified design utilises a meander-line resonator in the second layer while the reference design uses Helmholtz resonators except the second and thirtieth cells, which utilise three meander-line resonator. 
Their end distance $l$ is $5.5 mm$ and distance $d$ is $12 mm$. 

\subsection{Results}

The two 31-cell metalenses have been simulated in \textit{COMSOL}, excited by a normal incident plane-wave with pressure of 1 Pa. We first conduct simulations around the design frequency 3.0 kHz for both metalenses and the results are shown in Fig.~\ref{fig:4}a and Fig.~\ref{fig:4}b. The insets show the acoustic pressure field within the 2nd to 6th cells as these correspond to the critical impedance values that Helmholtz resonators fail to match. Exactly at the designed frequency 3.0 kHz, the focal intensity enhancement of the Helmholtz design (4.5) is higher than the modified design (3.5). However, it has been demonstrated in microwave metasurfaces\cite{olk_accurate_2019}, that resonant elements can have additional coupling through higher-order evanescent modes, which is neglected in a scalar impedance and can lead to a shift of the frequency of optimal performance. To give a fair comparison of performance, we show the intensity field of the modified design at 2990 Hz, 10 Hz less than the primary target frequency of 3.0 kHz. The modified design reaches enhancement of 5, which shows slight improvement over the Helmholtz design. To investigate the broadband performance of both structures, we perform a frequency sweep from 2.5 kHz to 3.2 kHz with a step of 5 Hz. To summarise the performance at each frequency, we consider the intensity profile along a vertical line through the center of each lens, with the results shown in Fig.~\ref{fig:5}a and b.

For the modified metalens, a clear focal spot can be observed between 0.25 m and 0.4 m from 2.6 kHz to nearly 3.0 kHz in the top panel in Fig.~\ref{fig:5}a. This deviates from our target focal length 0.4 m, which is due to the small size of our lens\cite{gao_analysis_2012}. Notably, some very strong resonances can be found around 2.7 kHz and 3.0 kHz above the metalens, due to Fabry-Perot resonances between the ends of the narrow duct. At these frequencies stored energy in the channel is very high, leading to strong near-field components near the metasurface observable in Fig.~\ref{fig:5}a. The modified design shows a broad focusing frequency band of over 400 Hz where it maintains a high intensity enhancement. We then choose frequencies corresponding to intensity peaks, indicated by dashed vertical lines in Fig.~\ref{fig:5}, and further investigate the intensity field patterns which are shown in the bottom panel in Fig.~\ref{fig:5}a. The focal spots show a good focusing pattern at these frequencies. Although the area of focal spots at 2765 Hz and 2905 Hz are larger than at the other two frequencies, their positions are very consistent as frequency varies. The top panel in Fig.~\ref{fig:5}b illustrates the plots for the Helmholtz metalens under the same conditions. We can find focusing zones from 2.7 kHz to 3.1 kHz, however, with much lower intensity enhancement than the modified design,  except for the region around 3.0 kHz. Comparable focusing effect can be realised around the design frequency 3.0 kHz, but this number decreases to 3, half of the peak, gradually after 2.95 kHz. We again pick four frequencies with good performance and inspect their intensity fields. It is clear that the Helmholtz based design has much weaker focal spot intensity at lower frequencies. However, the position of the focal spots remains consistent.

\begin{figure*}[t]
    \centering
    \subfigure{\includegraphics[width=0.25\textwidth]{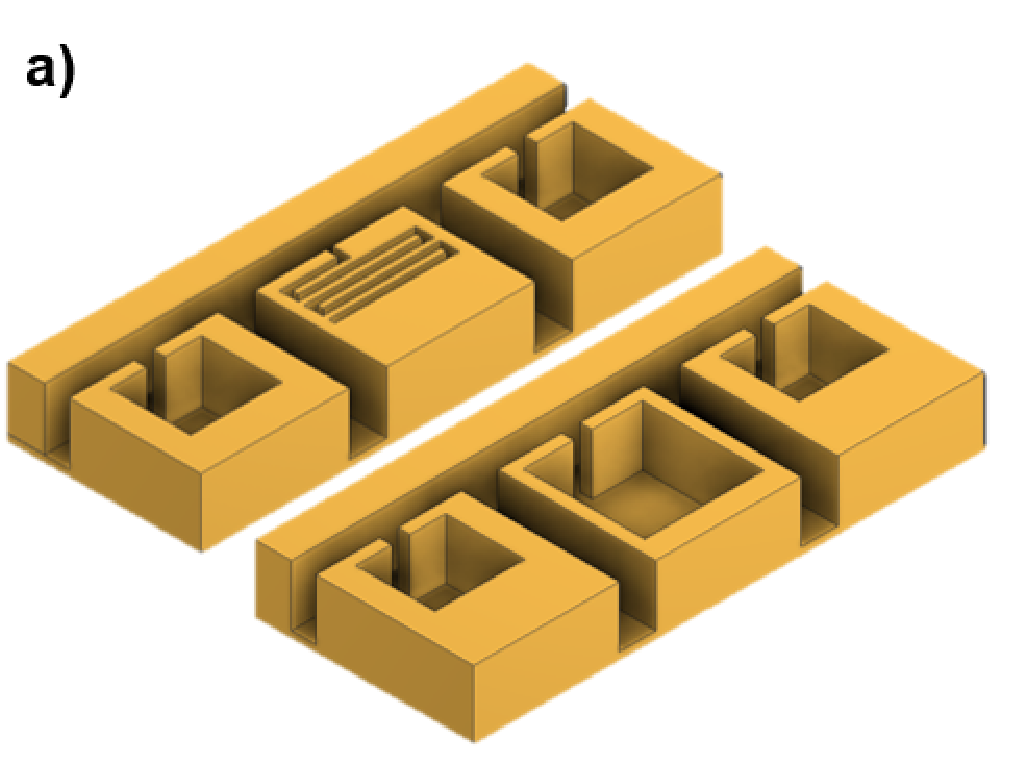}}
    \subfigure{\includegraphics[width=0.36\textwidth]{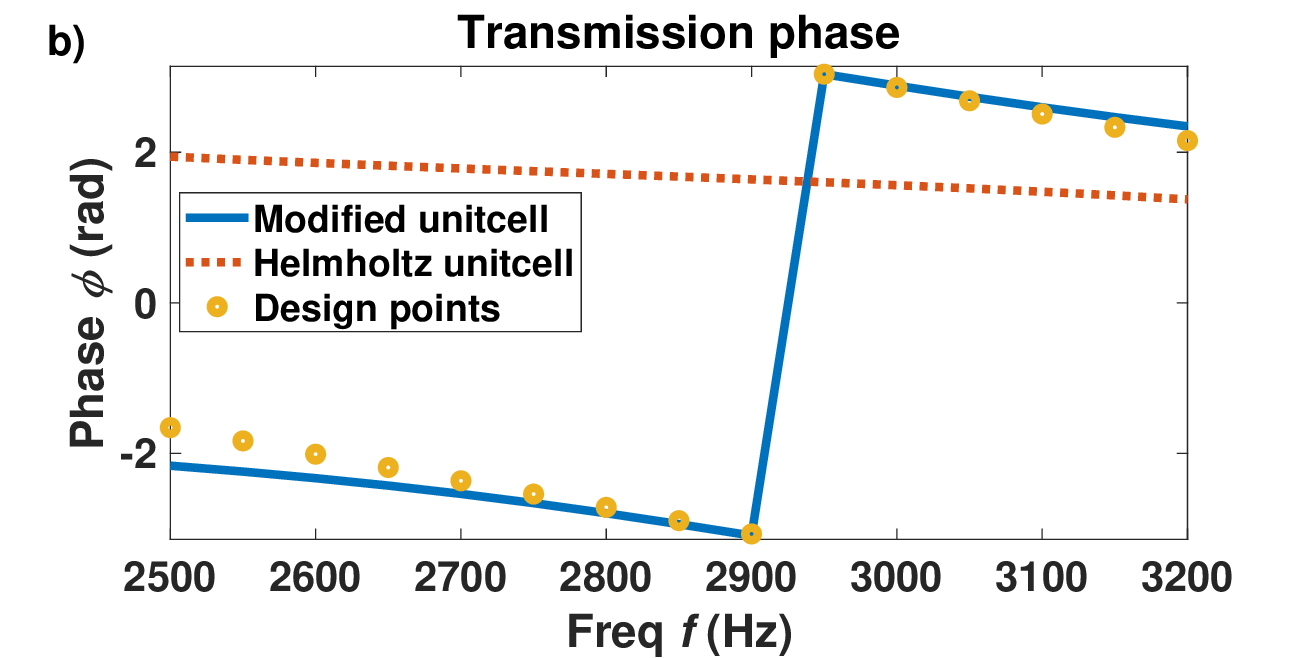}}
    \subfigure{\includegraphics[width=0.36\textwidth]{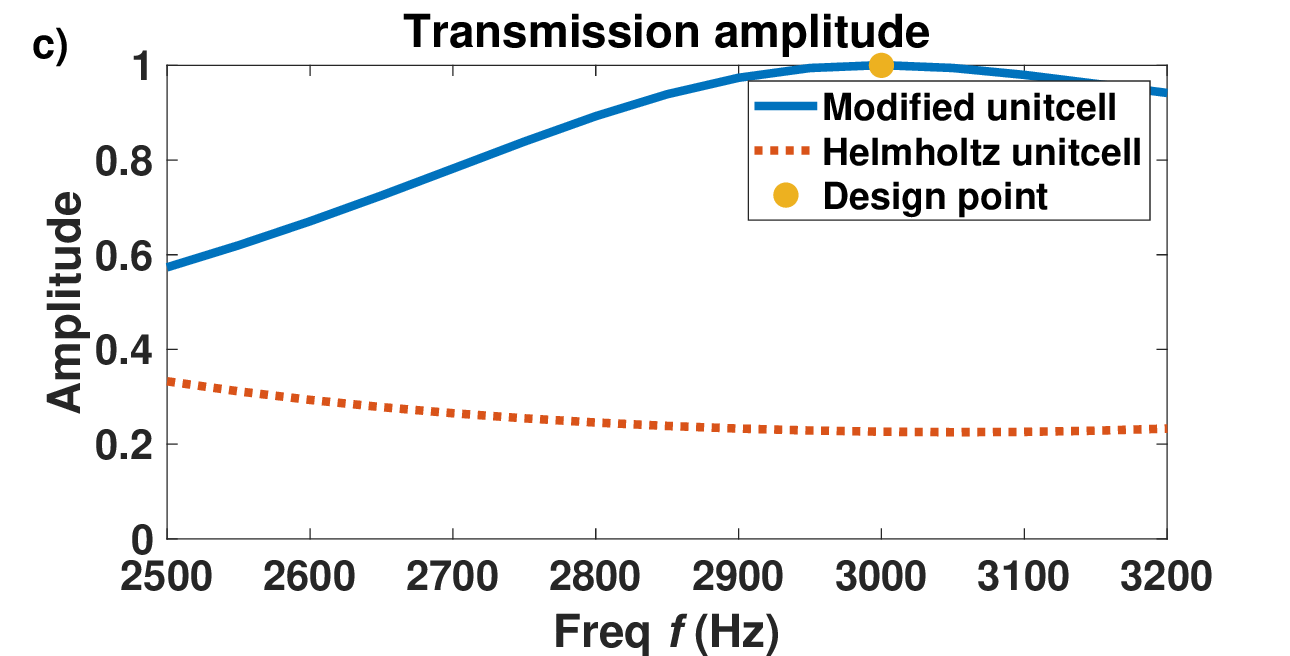}}
    \subfigure{\includegraphics[width=0.25\textwidth]{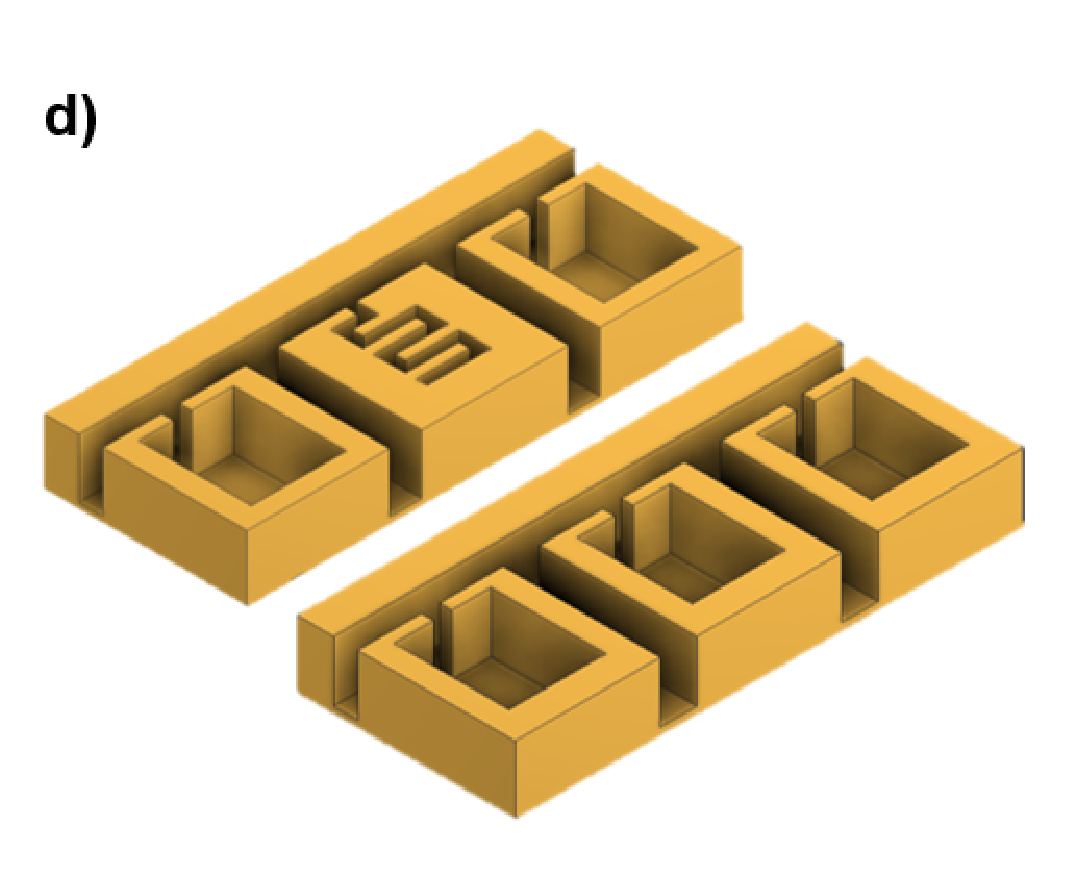}}
    \subfigure{\includegraphics[width=0.36\textwidth]{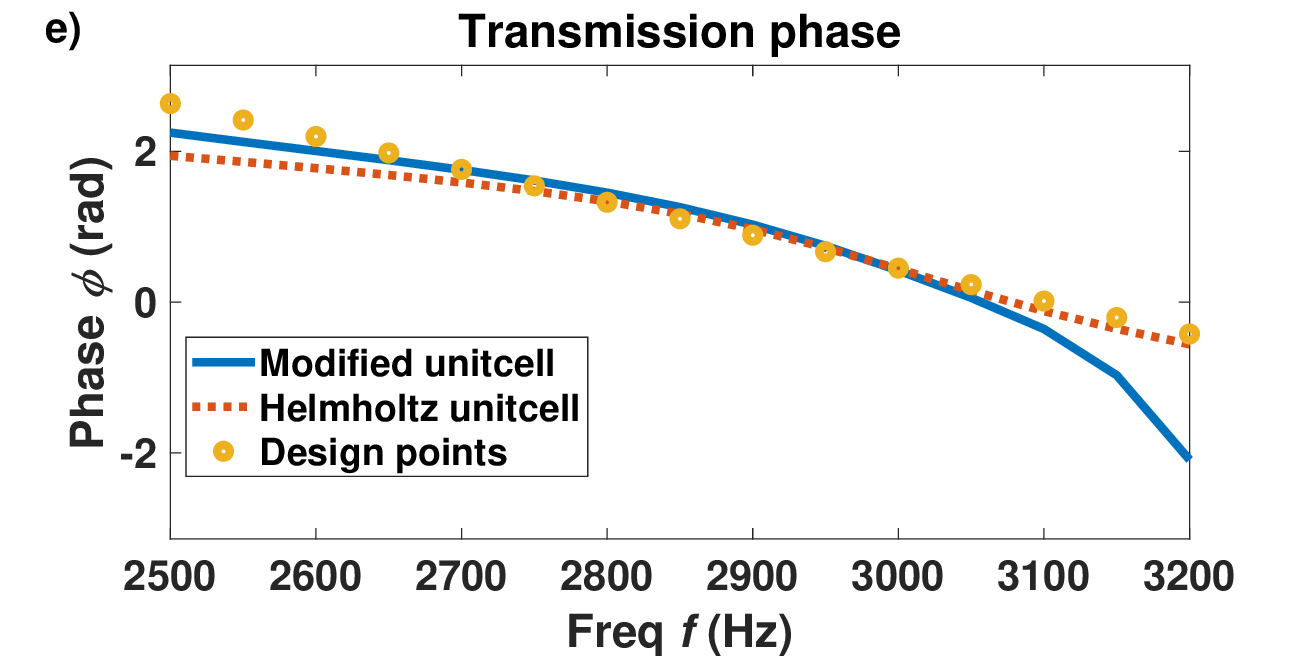}}
    \subfigure{\includegraphics[width=0.36\textwidth]{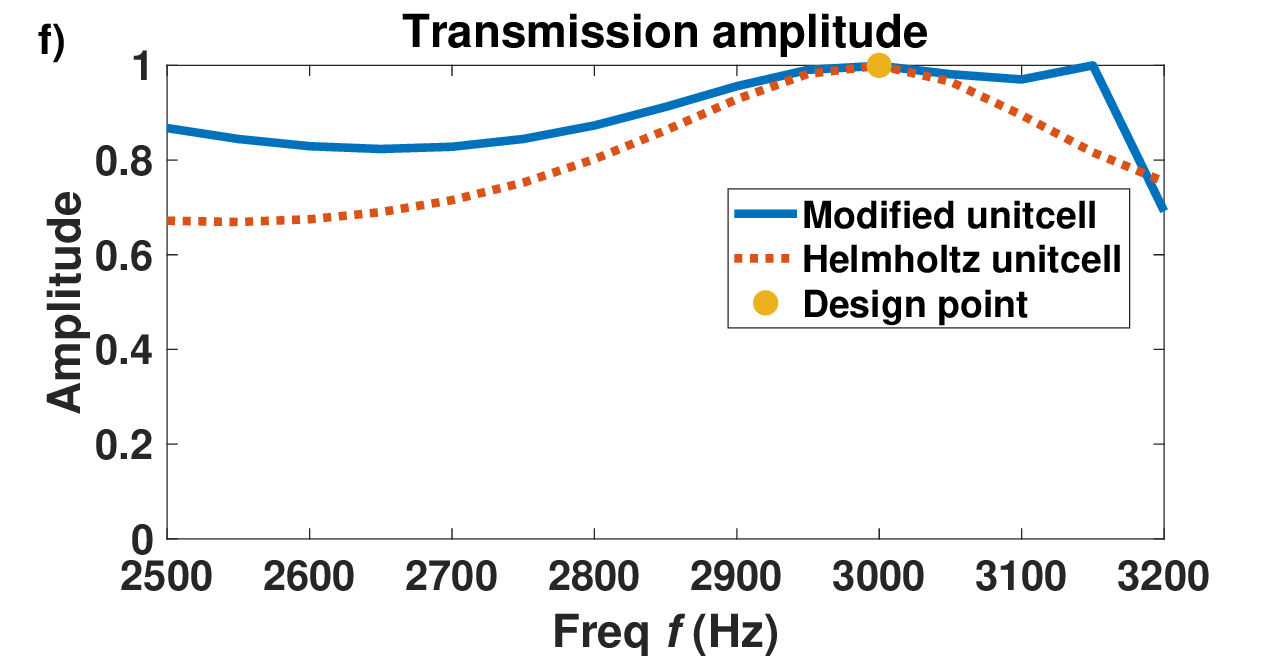}}

\caption{(a) The geometry of the third unit-cell in simulation: Top panel is the modified design and bottom panel is the Helmholtz design. (b) The transmission phase of these unit cells (c) The transmission amplitude of these unit cells (d) The geometry of the seventh unit-cell in simulation: Top panel is the modified design and bottom panel is the Helmholtz design. (e) The transmission phase of these unit cells (f) The transmission amplitude of these unit cells }
\label{fig:7}
 \end{figure*}

To identify the reason for the better broadband performance of the modified metalens, frequency sweep simulations have been conducted on the unit-cells of both metasurfaces. We consider two cases: the third unit-cell, where the Helmholtz resonators are unable to match the target impedance \(Z_2\), and the seventh unit-cell where the impedance is well matched by both designs. Fig.~\ref{fig:7} a and d show the geometries of third and seventh unit-cells of the design matching the phase profile in Fig.~\ref{fig:1}a. Fig.~\ref{fig:7}b,c,e,f show their transmission phase and amplitude, respectively. Unit-cell \#3 aims to obtain 2.83 rad with unity transmission. Referring to Fig.~\ref{fig:3}b, a significant difference can be found in \(Z_2\). A large discrepancy of the phase between the two unit-cells is obvious and the Helmholtz-only cell achieves a phase of 2 rads, 0.8 rad less than required while the modified cell matches well. Moreover, the amplitude of the Helmholtz design maintains a very low level near 0.37, much smaller than the modified one. Unit-cell \#7 targets a phase of 0.45 rad. The phase diagram shows that both of the unit-cells achieve the phase value at design frequency and their difference is very small from 2.5 kHz to 3.05 kHz. As for the amplitude, although they can both obtain near unity transmission at the design frequency 3.0 kHz, the amplitude of the modified cell is always higher than the Helmholtz design below 3.0 kHz. This scenario occurs in other unit-cells as well. Much lower amplitude and mismatched phases are discovered in Helmholtz unit-cells with great impedance mismatch. On the other hand, the phase values are close but the amplitude is still lower than the modified design for other well-matched impedances unit-cells. Consequently, the modified design can obtain a high intensity enhancement over a large frequency band as the unit-cells fit the phase better and have higher amplitude. 

\subsection{Influence of thermo-viscous losses}

\begin{figure*}[htb]
    \centering
    \subfigure{\includegraphics[width=0.48\textwidth]{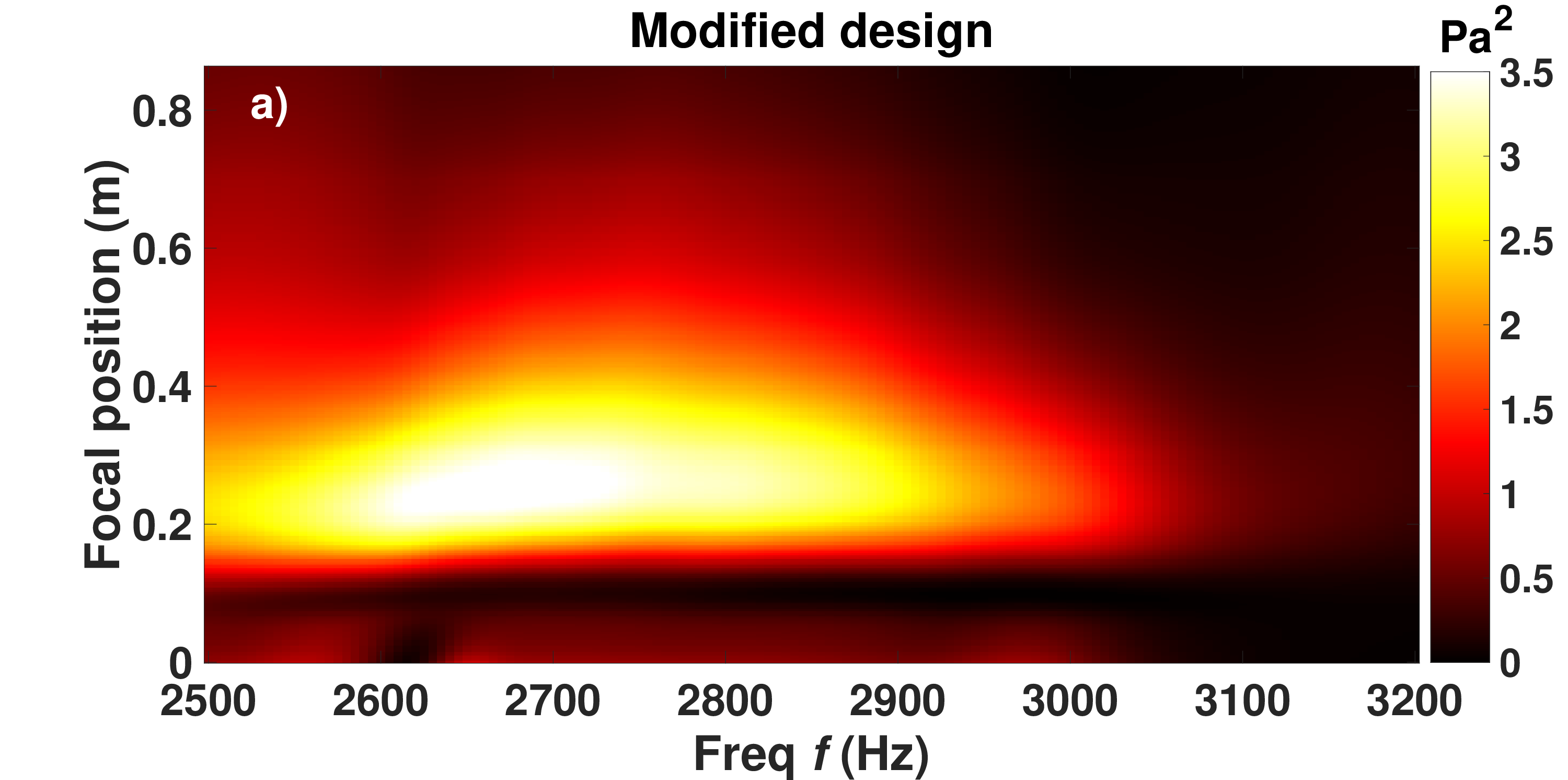}}
    \subfigure{\includegraphics[width=0.48\textwidth]{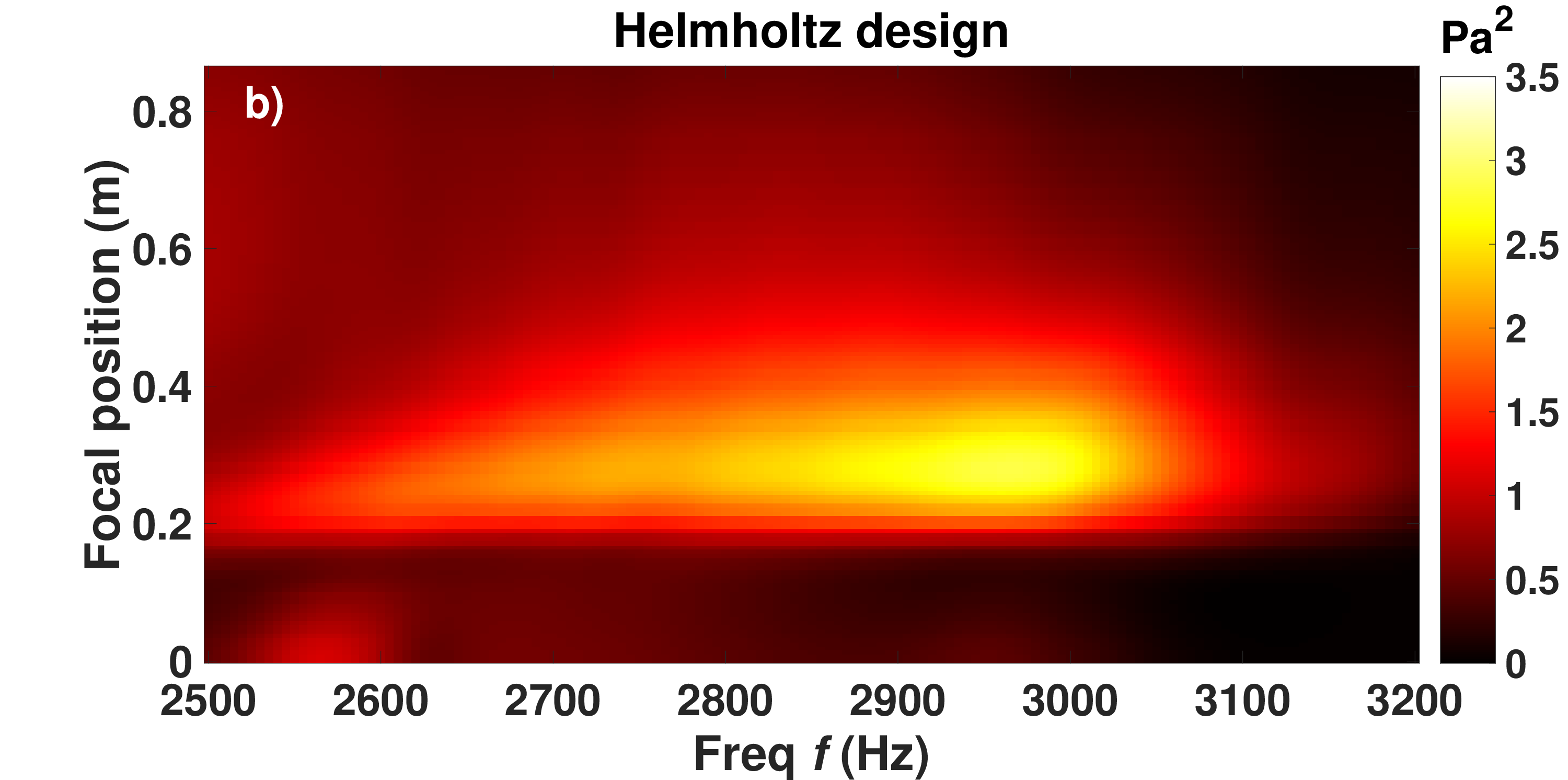}}
\caption{(a) The power intensity coefficient of the central line of the modified metalens in the frequency range from 2.5 kHz to 3.2 kHz with thermo-viscous loss. (b)The power intensity coefficient plot of the central line of the Helmholtz metalens in the frequency range from 2.5 kHz to 3.2 kHz with thermo-viscous loss.}
\label{fig:6}
 \end{figure*}

\begin{figure}[t]
    \centering
    \subfigure{\includegraphics[width=0.40\textwidth]{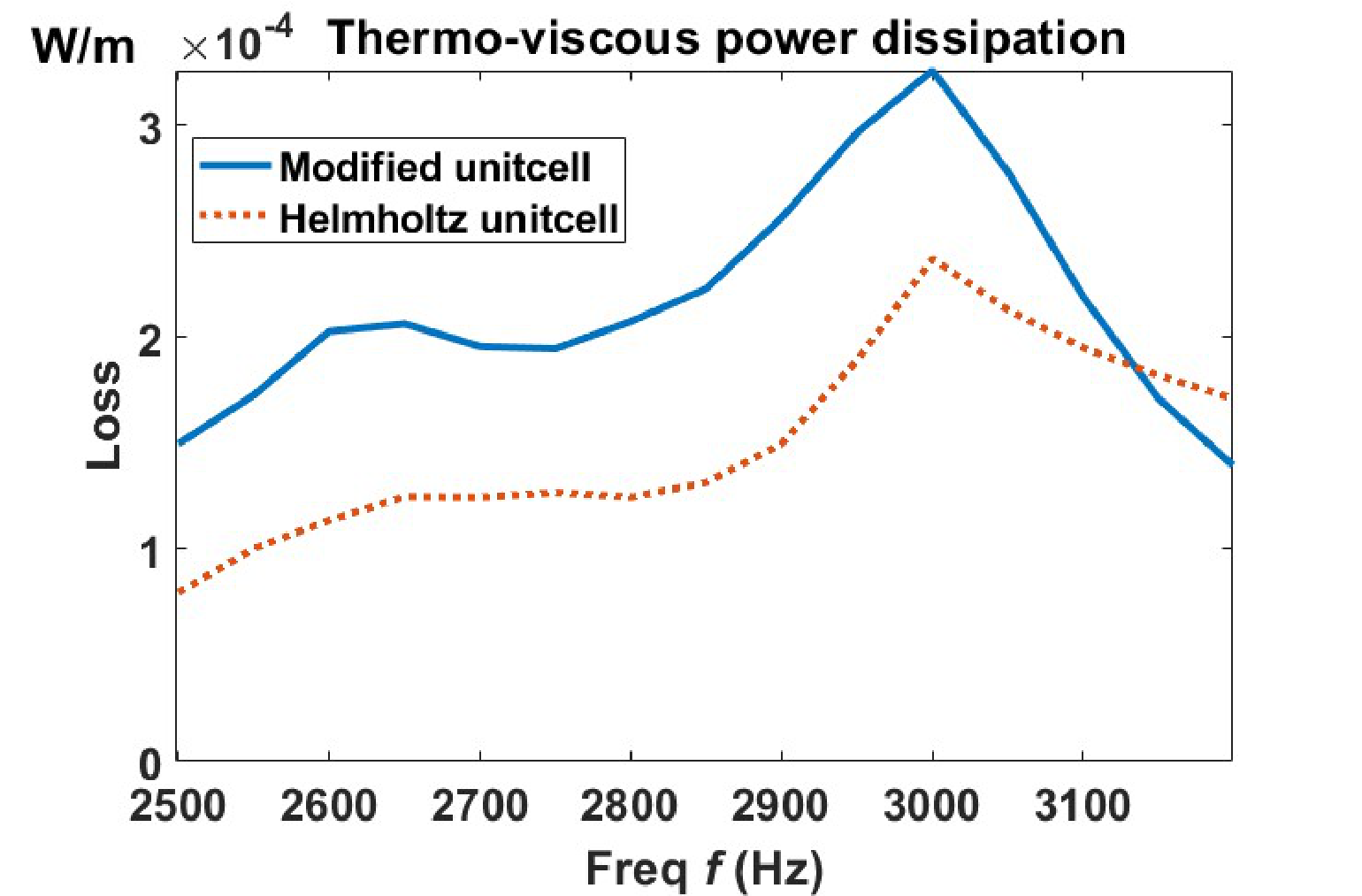}}

\caption{The thermo-viscous power dissipation}
\label{fig:loss}
 \end{figure}

Since acoustic propagation in narrow channels can lead to strong thermo-viscous losses, we conduct simulations to investigate how they influence the broadband focusing performance, with results plotted in Fig.~\ref{fig:6}. The effect of losses on power efficiency is significant, as both metalenses suffer a loss of 40\% of the peak value. For the metalens composed of three Helmholtz resonators, high intensity enhancement can be achieved at the design frequency of 3 kHz and gradually reduces as the frequency decreases. This corresponds to the trend in the lossless case in Fig.~\ref{fig:5}b. Although poor focusing is observed at 3 kHz for the modified metalens, a very similar zone appears from 2.5 kHz to 2.95 kHz. The thermo-viscous effects exert a downward shift in the best-performing region but the focusing bandwidth is consistent with the lossless case. The thermo-viscous power dissipation of both structures was calculated by integrating the loss density across the complete metasurface, with the results shown in Fig.~\ref{fig:loss}. The meander-line resonator suffers a greater 50\% energy loss than the Helmholtz cell on account of the winding narrower channels. Despite these higher losses, the meander-line structure maintains better focussing performance, due to its ability to cover the full range of required values, and its more favourable dispersion properties. We also note that the thermo-viscous losses dampen the peaks due to the Fabry-Perot resonances, making the focusing performance more consistent over the considered frequency range.

\section{Asymmetric unit cell design}
\label{sec:asymmetric}

\begin{figure*}[tb]
    \centering
    \subfigure{\includegraphics[width=0.23\textwidth]{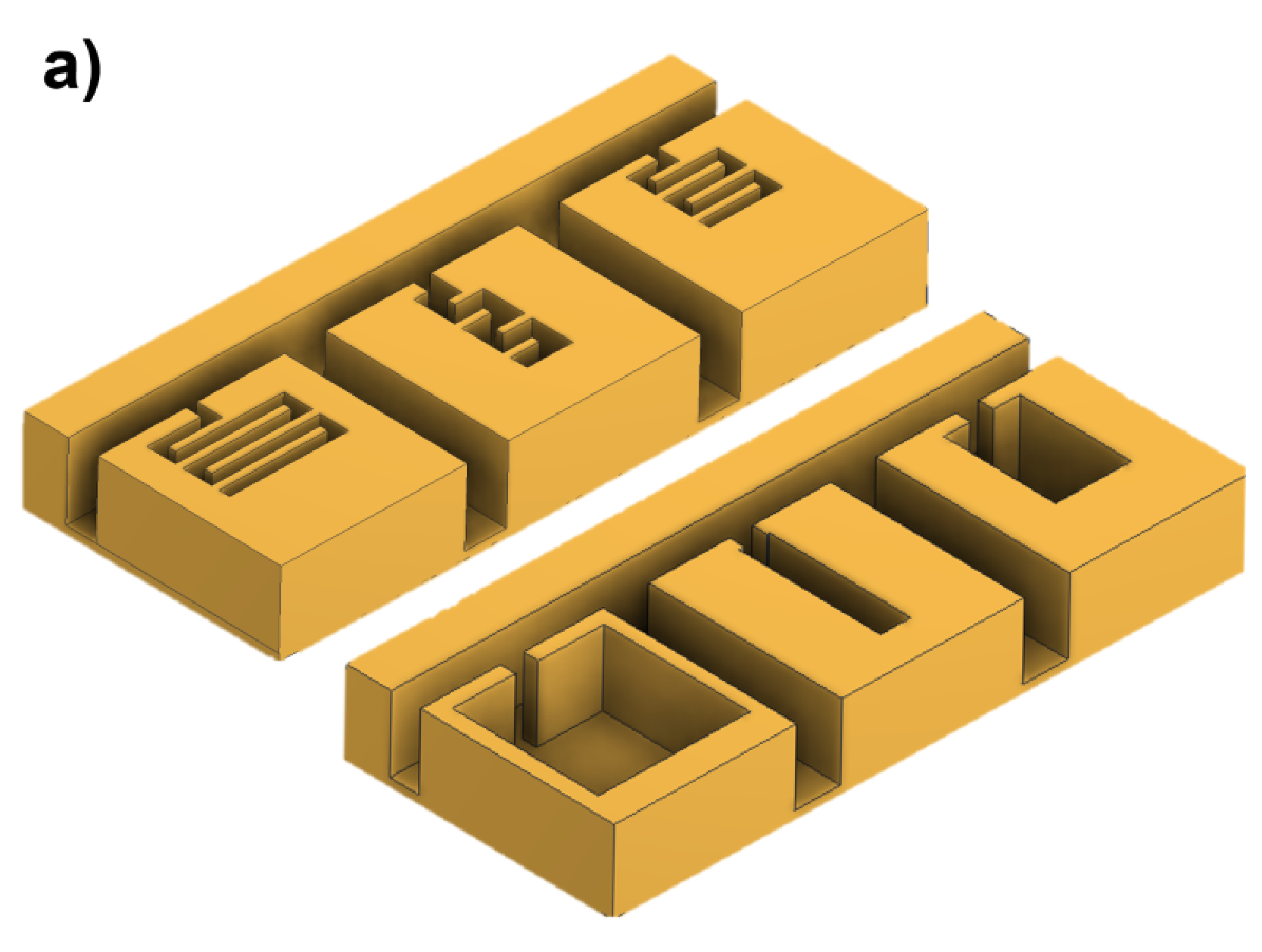}}
    \subfigure{\includegraphics[width=0.36\textwidth]{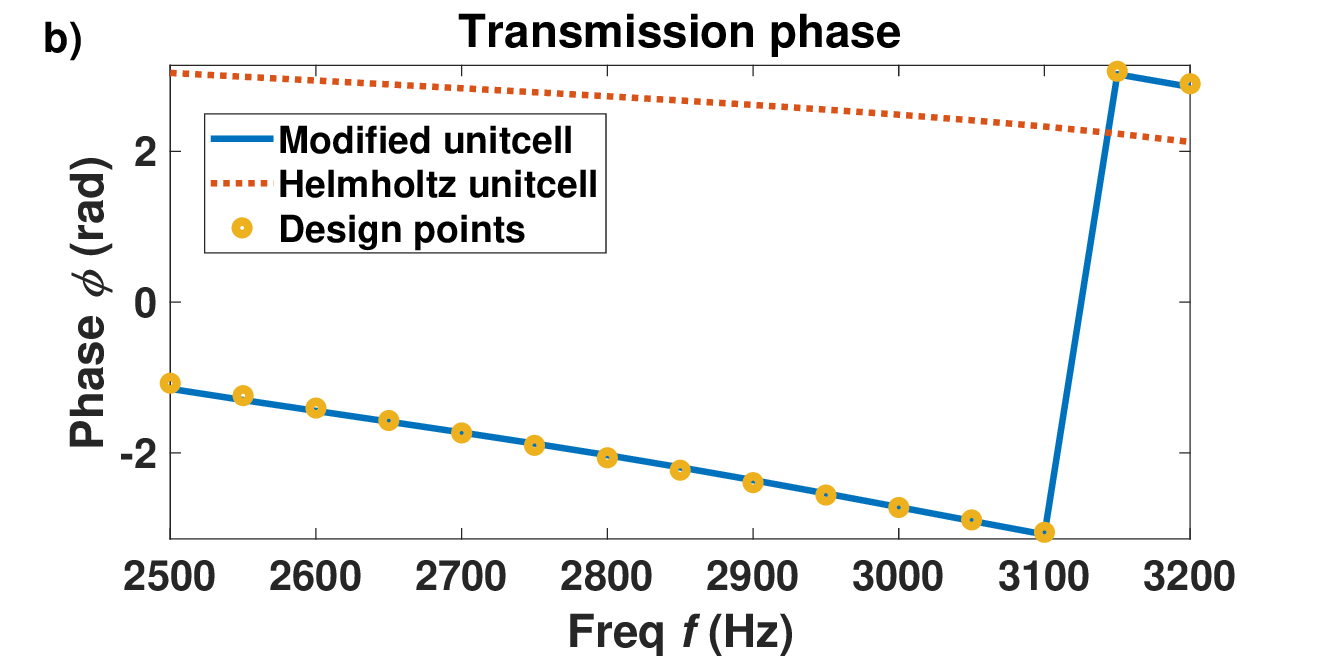}}
    \subfigure{\includegraphics[width=0.36\textwidth]{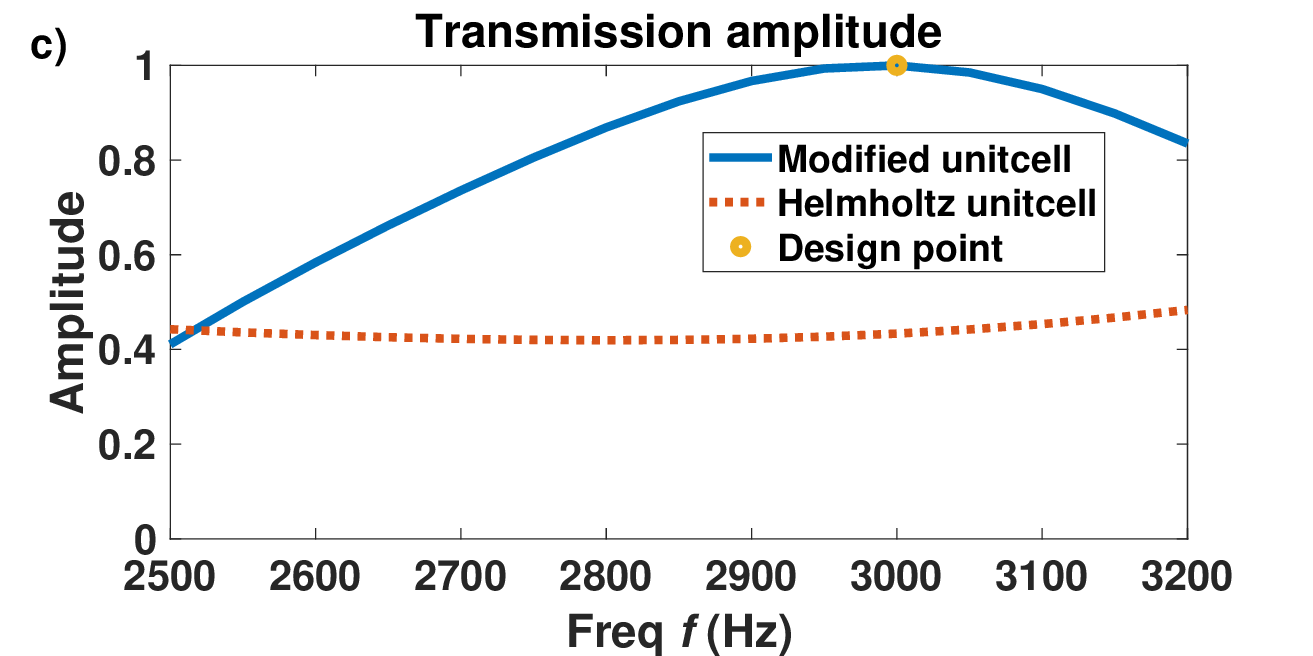}}

    \caption{(a) The geometry of the asymmetric unit-cell in simulation: Top panel is the modified design with three different meander-line resonators and bottom panel is the Helmholtz design. (b) The transmission phase of these unit cells (c) The transmission amplitude of these unit cells}
\label{fig:8}
 \end{figure*}

For lenses with high numerical aperture, and hence large refraction angle at the edges, the impedance of the refracted wave differs significantly from that of the designs. It is well-known\cite{huang_tunable_2024} that an asymmetric unit cell exhibiting bianisotropy can be impedance matched to both the incident and transmitted waves, enabling high transmission. The conversion among the two-port parameters has been modified to suit impedance change due to asymmetry. We obtain three different required reactance profiles in three positions receptively. We consider the case where the incident wave is normal and the transmitted wave should be refracted at an angle of 75 degrees. This leads to a wave reactance of \(2.755e4\) for the incident wave and \(1.065e5\) for the transmitted wave. Referring to Fig.~\ref{fig:2}b, the Helmholtz resonator cannot provide positive reactance. Nevertheless, two poles requiring large positive reactance can be found in the profiles of \(Z_1\) and \(Z_3\), suggesting that poor performance will be expected from Helmholtz unitcell due to impedance mismatch. On the other hand, the meander-line resonator can achieve most of the target phase thanks to its wider reactance coverage. We built two asymmetric unit-cells composed of all meander-line resonators and all Helmholtz resonators which are shown in Fig.~\ref{fig:8}a. We choose the critical case with a design phase of -2.72 rad because target \(Z_1\) is positive while \(Z_3\) is negative, indicating that only part of the Helmholtz resonators can meet the impedance profile. Their transmission phase and amplitude have been demonstrated in Fig.~\ref{fig:8}b and c. The modified cell achieves unity at the design frequency 3.0 kHz and much higher amplitude than Helmholtz design. It also obtains a perfect phase match with the exploited design points with group delay, exhibiting its promising potential in broadband applications. In summary, we extend our proposed method to solve the target impedance for an asymmetric case. Two unit-cells have been built and simulated. The results show the universality of our analytical model and the advantage of the proposed meander-line resonator over conventional Helmholtz resonator.

\section{\label{sec:conclusion}Conclusion}
In summary, we propose an integrated unit-cell for acoustic metasurfaces, containing both Helmholtz resonators and meander-lines. The meander-lines cell can approximate positive reactance better than the Helmholtz resonator within the space constraints of a metasurface. This enables unit cells with 3 resonators to be designed, having a closed-form solution which clearly shows how closely the design approahces the target transmission coefficient. The metalens has been designed and simulated, utilizing the presented meander-line structure. A reference metalens containing only Helmholtz resonators is also simulated for comparison. We find a stronger intensity enhancement (6.0) at the focal spot over the bandwidth of 400 Hz in the modified design. The lossy cases are also simulated and the modified metalens can still maintain focusing effect from 2.5 kHz to 2.9 kHz but the intensity enhancement reduces to 3.5. Then we show the frequency response for single unit-cells for both cases. The large mismatch between the Helmholtz resonator and the target impedance leads to phase shifts and great reflection and hence causes poor transmission efficiency. Lastly, we modified the analytical model for an asymmetric case. As more positive reactance values are required in the phase profile, our proposed meander-line resonator shows advantage of wider impedance coverage, resulting in a higher efficiency and excellent phase match compared to Helmholtz resonator. This work shows the unit-cell design from impedance engineering perspective and is promising to find applications in acoustic focusing, sensing and imaging.

\begin{acknowledgments}
This work was supported by Australian Research Council discovery project: DP200101708.
\end{acknowledgments}

\section{Reference}
\bibliography{MarRevisePaper}

\end{document}